\documentclass{article}
\usepackage[usenames,dvipsnames]{xcolor}
\usepackage{mathtools, amssymb, amsfonts, xargs, tensor, cite, stmaryrd, mathrsfs, graphicx, ifthen, pgfplotstable, tabularx, setspace, authblk, subcaption, Theorems}
\usepackage{pgfplots}
    \pgfplotsset{compat=1.18}
    \usepgfplotslibrary{statistics}
    
\DeclareMathOperator{\diag}{diag}
\DeclareMathOperator{\nullop}{Null}
\DeclareMathOperator{\spanop}{Span}
\DeclareMathOperator{\projop}{Proj}
\DeclareMathOperator{\rankop}{Rank}
\DeclareMathOperator{\rangeop}{Range}
\DeclareMathOperator{\uvecop}{uvec}
\DeclareMathOperator{\eigop}{eig}
\newcommand{\dif}{\mathop{}\mathrm{d}}
\begin{document}
    \title{Nonuniqueness and Convergence to Equivalent Solutions in Observer-based Inverse Reinforcement Learning
    \thanks{This research was supported, in part, by the National Science Foundation (NSF) under award numbers 1925147 and 2027999 and the Air Force Office of Scientific Research under award number FA9550-20-1-0127. Any opinions, findings, conclusions, or recommendations detailed in this article are those of the author(s), and do not necessarily reflect the views of the sponsoring agencies.}  }
    \author{Jared Town}
    \author{Zachary Morrison}
    \author{Rushikesh Kamalapurkar}
    \affil[1]{Oklahoma State University, Stillwater, OK, 74078, USA}
    \date{}
    \maketitle
    \begin{abstract}
        A key challenge in solving the deterministic inverse reinforcement learning (IRL) problem online and in real-time is the existence of multiple solutions. Nonuniqueness necessitates the study of the notion of equivalent solutions, i.e., solutions that result in a different cost functional but same feedback matrix, and convergence to such solutions. While \emph{offline} algorithms that result in convergence to equivalent solutions have been developed in the literature, online, real-time techniques that address nonuniqueness are not available. In this paper, a regularized history stack observer that converges to approximately equivalent solutions of the IRL problem is developed. Novel data-richness conditions are developed to facilitate the analysis and simulation results are provided to demonstrate the effectiveness of the developed technique.
    \end{abstract}
\section{Introduction}
    This paper concerns recovery of the cost functional being optimized by an \emph{expert} through observation of their input-output behavior. The expert is assumed to be controlling a deterministic dynamical system. The controller being implemented by the expert is assumed to be optimal with respect to an unknown cost functional. The objective of the learner is to estimate the cost functional using measurements of the experts inputs and outputs. Cost functional estimation techniques are studied in the literature under the umbrella of inverse reinforcement learning \cite{SCC.Ng.Russell2000}. While IRL typically includes utilization of the estimated cost functionals for behavior imitation using (forward) reinforcement learning, the scope of this paper is limited to cost functional estimation.
 
    IRL methods are often utilized to teach an autonomous system a specific task in an offline environment by observing repeated performance of the same task by the expert \cite{SCC.Ng.Russell2000,SCC.Russell1998,SCC.Abbeel.Ng2004,SCC.Ziebart.Maas.ea2008, SCC.Lian.Donge.ea2022, SCC.Self.Abudia.ea2022, SCC.Imani.Ghoreishi2022, SCC.Donge.Lian.ea2022, SCC.Inga.Bischoff.ea2019}. While effective, IRL techniques are generally offline, computationally complex, require multiple trajectories or several iterations over one trajectory, and require a greater amount of data than is readily available in real-time (online) applications. The aforementioned limitations are addressed in results such as \cite{SCC.Self.Coleman.ea2021a, SCC.Herman.Fischer.ea2015, SCC.Arora.Doshi.2017a} where online IRL methods that utilize a single iteration over one continuous trajectory are developed to learn the cost functional of the expert. New techniques to solve the IRL problem up to a scaling factor through non-cooperative linear quadratic differential games are also developed in \cite{SCC.Inga.Bischoff.ea2019} and \cite{SCC.Inga.Creutz.ea2021}.

    Results such as \cite{SCC.Self.Coleman.ea2021a, SCC.Herman.Fischer.ea2015, SCC.Arora.Doshi.2017a, SCC.Inga.Creutz.ea2021} (implicitly or explicitly) assume that the IRL problem admits a unique solution. Since IRL problems generally admit multiple linearly independent solutions \cite{SCC.Jameson.Kreindler1973, SCC.Jean.Maslovskaya2018}, the uniqueness assumption is restrictive. Non-uniqueness is studied in results such as \cite{SCC.Jameson.Kreindler1973}, where procedures to determine equivalent cost functionals are developed. It is also shown that IRL problems with multiple solutions arise naturally in state space models that have a product structure (see \cite{SCC.Jean.Maslovskaya2018}). Many real-world systems have a product structure, either in the original model or in the linearized model. For example, linearized dynamics of aerospace vehicles have a product structure due to separation of longitudinal and lateral dynamics \cite{SCC.Jean.Maslovskaya2018}. The study of IRL problems that admit multiple solutions is thus indispensable in real-world applications.
    
    The IRL methods recently developed in results such as \cite{SCC.Lian.Xue.ea2021a, SCC.Donge.Lian.ea2022, SCC.Xue.Kolaric.ea2021} study nonuniqueness of solutions to IRL problems and guarantee convergence to the set of equivalent solutions. In \cite{SCC.Donge.Lian.ea2022, SCC.Xue.Kolaric.ea2021} the IRL problem is solved in an offline setting as opposed to the online and real-time problem under consideration in this paper. In results such as \cite{SCC.Donge.Lian.ea2022, SCC.Lian.Xue.ea2021a} equivalent solutions for the state penalty matrix are identified, using measurements of only the control input of the expert. However, these results do not estimate the control penalty of the expert. The technique developed in this paper requires more information than \cite{SCC.Donge.Lian.ea2022, SCC.Lian.Xue.ea2021a} (measurements of the control input \emph{and the output} of the expert), but in contrast with \cite{SCC.Donge.Lian.ea2022, SCC.Lian.Xue.ea2021a}, the entire cost functional of the expert, including state \emph{and control} penalties, is estimated.
         
    Motivated by \cite{SCC.Self.Coleman.ea2021a}, the method developed in this paper identifies an equivalent cost functional for the expert given measurements of the control input and the output of the expert in an observer framework. Specifically, the History Stack Observer (HSO) from \cite{SCC.Self.Coleman.ea2021a}, originally designed under the uniqueness assumption, is extended to IRL problems that admit multiple solutions. The re-designed HSO is a true extension of the HSO from \cite{SCC.Self.Coleman.ea2021a} in the sense that it identifies the true cost functional of the expert, up to a scaling factor, if the IRL problem has a unique solution. While nonuniqueness is studied in the observer context in \cite{SCC.Town.Morrison.ea2023}, the definition of equivalence used in this paper is stronger than the one in \cite{SCC.Town.Morrison.ea2023}. As a result, the analysis that proves convergence to equivalent solutions is more involved than the analysis in \cite{SCC.Town.Morrison.ea2023}. In addition, the practically relevant case of convergence to approximately equivalent solutions is studied in this paper.

    This article extends the IRL HSO in \cite{SCC.Self.Coleman.ea2021a} to problems where the observed trajectories can be optimal with respect to multiple cost functionals. A learner with access to the state space model, controller input, and measurement data reconstructs an equivalent cost functional of an expert. Since recovery of the true cost functional cannot be expected in such problems, analysis of the error between the estimated cost functional and the true cost functional, as done in \cite{SCC.Self.Coleman.ea2021a}, is no longer useful. In this paper, a novel analysis approach that guarantees convergence of the learned solution to a neighborhood of an equivalent solution is developed. Under sufficient data informativity conditions, a new equivalence metric is designed such that convergence of the equivalence metric to zero implies convergence to an equivalent solution. The developed modification to the HSO is inspired by ridge regression, but has a surprising convergence property. Under ideal conditions (no noise and persistently exciting regressor), the convergence is exact, as opposed to ridge regression, where the solutions are off by a factor proportional to the regularization coefficient.
    	
\section{Problem Formulation}\label{Section: Probs Form}
The system being controlled by the expert is assumed to be a linear system of the form
\begin{equation}\label{Equation: Linear System}
    \dot{x}(t) = Ax+Bu,
\end{equation}
with output
\begin{equation}
    y = Cx(t),
\end{equation}
where the state is
$x\in\mathbb{R}^{n}$ and the control input is $u\in\mathbb{R}^{m}$. The system matrices are given as $A \in \mathbb{R}^{n\times n}$ and $B \in \mathbb{R}^{n\times m}$, and the output and output matrix are given as $y \in \mathbb{R}^{L}$ and $C \in \mathbb{R}^{L\times n}$ respectively. 		

The expert is assumed to implement an optimal controller that optimizes the cost functional
\begin{equation}\label{cost_fun}
        J(x_0,u({\cdot}))=
        \int_{0}^{\infty} \left(x(t)^{\top}Qx(t) + u(t)^{\top}Ru(t)\right)\dif t, 
\end{equation}
where $x(\cdot)$ is the system trajectory under the optimal control signal $u(\cdot)$ and starting from the initial condition $x_0$, $Q \in\mathbb{R}^{n\times n}$ is an unknown positive semi-definite matrix, and $R\in\mathbb{R}^{m\times m}$ is an unknown positive definite matrix. The following assumption ensures that the IRL problem is well-posed.       
\begin{assumption}\label{Assumption: Well-posed}
    The pair $(A,B)$ is stabilizable and the pairs $(A,C)$ and $(A,\sqrt{Q})$ are detectable.
\end{assumption}
Stabilizability of $(A,B)$ and detectability of $(A,\sqrt{Q})$ is needed for the optimal controller to exist and detectability of $(A,C)$ guarantees the existence of a matrix $L$ such that $A-LC$ is Hurwitz \cite[Lemma 21.1]{SCC.Hespanha2009}. Under Assumption \ref{Assumption: Well-posed}, the policy of the expert is given by $u = K_{Ep}x$, where $K_{Ep}\in\mathbb{R}^{m \times n}$ is obtained by solving the algebraic Riccati equation (ARE) corresponding to the optimal control problem described by the system in \eqref{Equation: Linear System}  and the cost functional in \eqref{cost_fun}.  

The learning objective is to estimate, online and in real-time, the unknown matrices in the cost functional using knowledge of the system matrices, $A$, $B$, and $C$, and input-output data. Generally, for a system $(A,B,C)$, a given set of input-output trajectories is optimal with respect to multiple cost functionals. As a result, the true cost functional cannot generally be estimated from data. Instead, an equivalent solution to the IRL problem is sought (see Definition \ref{Definition: Equivalent Solution} and \cite{SCC.Xue.Kolaric.ea2021}).

While the HSO in \cite{SCC.Self.Coleman.ea2021a} is an effective technique to solve the IRL problem online and in real-time, the analysis focuses on the error between the true cost functional matrices and their estimates, and as such, implicitly assumes uniqueness of solutions. As such, the method in \cite{SCC.Self.Coleman.ea2021a} cannot be applied to a large class of IRL problems that admit multiple solutions. In this paper, the HSO is extended to be applicable to IRL problems that admit multiple solutions. While the extension is similar to the regularization used in ridge regression, the fact that the error between the true cost functional matrices and the obtained estimates can no longer be used as a metric to gauge quality of the estimates necessitates the development of a novel analysis approach.

\section{Nonuniqueness and the History Stack Observer}\label{Section: HSO_Mods}
To facilitate the discussion, this section provides a brief summary of the HSO developed in \cite{SCC.Self.Coleman.ea2021a} and highlights the key problem that is resolved in this paper.
\subsection{Equivalent Solutions and Equivalence Metric}
If the state and control trajectories of the system are optimal with respect to the cost functional in \eqref{cost_fun} and Assumption \ref{Assumption: Well-posed} is met, then there exists a matrix $S$ such that for all $t\geq 0$, the matrices $Q$, $R$, $A$, and $B$, and the optimal trajectories $x(\cdot)$ and $u(\cdot)$ satisfy the Hamilton-Jacobi-Bellman (HJB) equation
\begin{equation}\label{HJB}
    x^{\top}(t)\left(A^{\top}S+SA-SBR^{-1}B^{\top}S+Q\right)x(t)=0,
\end{equation}
and the optimal control equation
\begin{equation}\label{OptCTL}
    u(t) = u^*(x(t)) \coloneqq -R^{-1}B^{\top}Sx(t).
\end{equation}
The feedback matrix of the expert is then given by $K_{Ep} = R^{-1}B^{\top}S$. The HJB equation and the optimal control equation facilitate the definition of an equivalent solution.
\begin{definition}\label{Definition: Equivalent Solution}
A solution ($\hat{Q}$, $\hat{S}$, $\hat{R}$) is called an equivalent solution of the IRL problem if it satisfies the ARE $A^{\top}\hat S+\hat S A- \hat SB\hat R^{-1}B^{\top}\hat S+\hat Q=0$ and optimization of the performance index $J$, with $Q=\hat{Q}$ and $R=\hat{R}$, results in the same feedback matrix as the one utilized by the expert, that is, $\hat{K}_P \coloneqq \hat R^{-1}B^{\top} \hat S =  K_{Ep}$.
\end{definition}
Given an estimate $\hat x$ of the state $x$, a measurement of the control signal, $u$, and estimates $\hat Q$, $\hat R$, and $\hat S$ of $Q$, $R$, and $S$, respectively, \eqref{HJB} and \eqref{OptCTL} can be evaluated to develop an observation error that evaluates to zero if the state estimates are correct and ($\hat Q$, $\hat R$, $\hat S$) is an equivalent solution. The observation error is then used to improve the estimates by framing the IRL problem as a state estimation problem. The rest of this subsection is borrowed from \cite{SCC.Self.Coleman.ea2021a} and is included here for completeness.

To facilitate the observer design, equations \eqref{HJB} and \eqref{OptCTL}
are linearly parameterized as 
\begin{align}
    0&=2\sigma_{R2}(u)W_{R}^{*} + B^{\top}\left( \nabla_{x}\sigma_S(x)\right)^{\top}W_{S}^{*}, \label{CRE_0}\\
    0&=\nabla_x\left((W_{S}^{*})^{\top}\sigma_S(x)\right)\left(Ax+Bu\right)
     +(W_{Q}^{*})^{\top}\sigma_Q(x)+(W_{R}^{*})^{\top}\sigma_{R1}(u),\label{IBE_0}
\end{align}
where $ x^{\top}Sx=(W_{S}^{*})^{\top}\sigma_S(x)$, $x^{\top}Qx=(W_{Q}^{*})^{\top}\sigma_Q(x)$, $u^{\top}Ru = (W_{R}^{*})^{\top}\sigma_{R1}(u)$, and $Ru = \sigma_{R2}(u)W_{R}^{*}$, where $\left[W_S^{*\top},W_Q^{*\top},W_R^{*\top}\right]^{\top}$ $\in\mathbb{R}^{P_S}\times\mathbb{R}^{P_Q}\times\mathbb{R}^{M}$ are the ideal weights with $P_S$, $P_Q$, and $M$ being the number of basis functions in the respective linear parameterization.  The ideal weights are given by
\begin{align*}
    W_S^* &= \left[ S_{11},S_1^{(-1)}, S_{22}, S_2^{(-2)},\ldots, S_{n-1}^{-(n-1)}, S_{nn}\right]^{\top} \\    
    W_Q^* &= \left[ Q_{11},Q_1^{(-1)}, Q_{22}, Q_2^{(-2)},\ldots, Q_{n-1}^{-(n-1)}, Q_{nn}\right]^{\top} \\  
    W_R^* &= \left[ R_{11},R_1^{(-1)}, R_{22}, R_2^{(-2)},\ldots, R_{m-1}^{-(m-1)}, R_{nn}\right]^{\top}
\end{align*}
The basis functions are given by 
\begin{align*}
    &\sigma_S(x)=\sigma_Q(x):=[x_1^2,2x_1x_2,2x_1x_3,\ldots,2x_1x_n,x_2^2,2x_2x_3,2x_2x_4,\ldots,x_{n-1}^2,\ldots,\nonumber\\
    &\qquad\qquad 2x_{n-1}x_{n},x_{n}^2]^\top,\nonumber\\
    &\sigma_{R1}(u):=[u_1^2,2u_1u_2,2u_1u_3,\ldots,2u_1u_m, u_2^2,2u_2u_3,2u_2u_4,\ldots,u_{m-1}^2,\ldots,\nonumber\\
    &\qquad\qquad 2u_{m-1}u_{m},u_{m}^2]^\top,
\end{align*}
and 
\begin{equation}
            \sigma_{R 2}(u)=
            \begin{bmatrix}
                u^{\top} & 0_{1 \times m-1} & 0_{1 \times m-2} & \ldots & 0 \\
                u_{1}e_{2}^{m} & \left(u^{\top}\right)^{(-1)} & 0_{1 \times m-2} & \ldots & 0 \\
                u_{1}e_{3}^{m} & u_{2}e_{2}^{m-1} & \left(u^{\top}\right)^{(-2)} & \ldots & 0 \\
                \vdots & \vdots& \vdots & \ddots & \vdots \\
                u_{1}e_{m}^{m} & u_{2}e_{m-1}^{m-1} & u_{3}e_{m-2}^{m-2} & \cdots & u_{m}
            \end{bmatrix}.\label{eq:u_basis}
\end{equation}
In \eqref{eq:u_basis}, $u^{(-j)}$ denotes the vector $u$ with the first $j$ elements removed, $e_{i}^{j}$ denotes a row vector of size $j$, with a one in the $i-$th position and zeros everywhere else.

Using the estimates $\hat{W}_S$, $\hat{W}_Q$, and $\hat{W}_R$ for $W_S^*$, $W_Q^*$, and $W_R^*$, respectively, in \eqref{CRE_0} and \eqref{IBE_0}, a control residual error and an inverse Bellman error are defined as
\begin{align}
    \Delta_u^\prime &\coloneqq 2\sigma_{R2}(u)\hat{W}_{R} + B^{\top}\left( \nabla_{x}\sigma_S(x)\right)^{\top}\hat{W}_{S}\text{ and}\label{CRE}\\ 
     \delta^\prime &\coloneqq \nabla_x\left((\hat{W}_{S})^{\top}\sigma_S(x)\right)\left(Ax+Bu\right) +(\hat{W}_{Q})^{\top}\sigma_Q(x)+(\hat{W}_{R})^{\top}\sigma_{R1}(u)\label{IBE}.
\end{align}

The scaling ambiguity inherent in linear quadratic optimal control, which is apparent in the fact that $\hat{W}^\prime= [\hat{W}_S^\top, \hat{W}_Q^\top, \hat{W}_{R}^{\top}]^\top = 0$ is a solution of \eqref{CRE_0} and \eqref{IBE_0}, is resolved, without loss of generality, by assigning an arbitrary value to one element of $\hat{W}^\prime$. Selecting the first component of $\hat{W}_R$ to be equal to $r_1 > 0$ and removing it from the weight vector $\hat{W}^\prime$ in \eqref{CRE_0} and \eqref{IBE_0} yields scale-aware definitions of the control residual error and the inverse Bellman error, given by
\begin{equation}\label{IBE_CRE}
    \begin{bmatrix}
        \delta\left(x, u, \hat{W}\right)\\
        \Delta_u\left(x, u, \hat{W}\right)
    \end{bmatrix}
    =
    \begin{bmatrix}
        \sigma_\delta(x,u)\\
        \sigma_{\Delta_u}(x,u)
    \end{bmatrix}
    \begin{bmatrix}
        \hat{W}_S \\ \hat{W}_Q \\ \hat{W}_R^{-}
    \end{bmatrix}
    +
    \begin{bmatrix}
        u_1^2r_1 \\ 2u_1r_1 \\ 0_{m-1\times 1}
    \end{bmatrix},
\end{equation}
where $\hat{W}_R^-$ is a copy of  $\hat{W}_R$ with the first element removed, $\sigma_\delta$ is a copy of $\big[(Ax+Bu)^{\top}(\nabla_x\sigma_S(x))^{\top},\allowbreak \sigma_Q(x)^{\top}, \sigma_{R1}(u)^{\top}\big]$, with the $(P_S + P_Q + 1)-$th element removed, and $\sigma_{\Delta_u}$ is a copy of $\big[B^{\top}(\nabla_x\sigma_S(x))^{\top},\allowbreak 0_{m\times P_Q}, 2\sigma_{R2}(u)\big]$, with the $(P_S + P_Q + 1)-$th column removed. In this paper, the error system in \eqref{IBE_CRE} is used as an \emph{equivalence metric} to develop an observer-based IRL method. The following section provides a brief overview of the observer developed in \cite{SCC.Self.Coleman.ea2021a}.

\subsection{The History Stack Observer \label{subsec:HSO}}
Pairing the innovation $y - C\hat x$ with the inverse bellman error and control residual error from \eqref{IBE_CRE} yields the observation error $\omega= \begin{bmatrix}Cx\\ \Sigma_u \end{bmatrix} - \begin{bmatrix}C\hat{x}\\ \hat{\Sigma}\hat{W}\end{bmatrix}$, where $\hat{W} = [\hat{W}_S^\top, \hat{W}_Q^\top, (\hat{W}_{R}^{-})^{\top}]^\top$,
\begin{gather*}
    \hat{\Sigma} \coloneqq
    \begin{bmatrix} \sigma_\delta\left(\hat{x}(t_1),u(t_1)\right)\\ \sigma_{\Delta_u}\left(\hat{x}(t_1),u(t_1)\right)\\
    \vdots\\
    \sigma_\delta\left(\hat{x}(t_N),u(t_N)\right)\\ \sigma_{\Delta_u}\left(\hat{x}(t_N),u(t_N)\right)
    \end{bmatrix}, \text{ and }
    \Sigma_u \coloneqq
    \begin{bmatrix}
    -u_1^2(t_1)r_1\\
    -2u_1(t_1)r_1\\
    0_{m-1\times 1}
    \\
    \vdots\\
    -u_1^2(t_N)r_1\\
    -2u_1(t_N)r_1\\
    0_{m-1\times 1}
    \end{bmatrix},
\end{gather*}
Using the observation error, the history stack observer is designed in \cite{SCC.Self.Coleman.ea2021a} as
\begin{equation}\label{Hopefully_is_update_law}
    \begin{bmatrix}
        \dot{\hat{x}}\\ \dot{\hat{W}}
    \end{bmatrix} = 
    \begin{bmatrix}
        A\hat{x}+Bu\\ 0_{P_S + P_Q + M-1}\end{bmatrix} + K
        \left(\begin{bmatrix}
                Cx\\ \Sigma_u
            \end{bmatrix} - 
            \begin{bmatrix}
                C\hat{x}\\ \hat{\Sigma}\hat{W}
            \end{bmatrix}\right),
\end{equation}
where the gain $K$ is selected as
\begin{equation}\label{Equation: K HSO}
    K \coloneqq
    \begin{bmatrix}
            K_{3} & 0_{n\times N+Nm}\\
            0_{P_S+P_Q+M-1\times L} & K_4(\hat{\Sigma}^{\top}\hat{\Sigma})^{-1}\hat{\Sigma}^{\top}
    \end{bmatrix},
\end{equation}
where $K_3$ is selected so that $A - K_3C$ is Hurwitz, and $K_4$ is scalar multiple of an identity matrix of size $P_S+P_Q+M-1$. To facilitate the analysis, let $\Sigma$ be a copy of $\hat{\Sigma}$ where the state estimates are replaced by their true values and let $W^*\coloneqq (r_1/W_R^{*}(1))[W_S^{*\top},W_Q^{*\top},(W_{R}^{-*})^{\top}]^\top$, where $W_R^{-*}$ denotes $W_R^{*}$ with the first element, $W_R^{*}(1)$, removed.

The matrices $\hat \Sigma\in \mathbb{R}^{N(m+1)\times P_S+P_Q+M-1}$ and $\Sigma_u\in \mathbb{R}^{N(m+1)}$ are constructed using the dataset $\{(\hat x(t_i),u(t_i))\}_{i=1}^{N}$, recorded at time instances $\{t_1, \ldots t_N\}$, with $N \ge P_S + P_Q + M - 1$. The dataset is referred to hereafter as a \emph{history stack}. To ensure convergence of the weights, updated using \eqref{Hopefully_is_update_law}, to an equivalent solution (see Theorem \ref{Theorem: Delta Convergence} below), the history stack is recorded using a condition number minimization algorithm. At any time, two separate history stacks, $H_1$ and $H_2$ are maintained. The history stack $H_1$ is used to compute the matrices $\hat \Sigma$ and $\Sigma_u$ in \eqref{Hopefully_is_update_law} and $H_2$ is populated with current state estimates and control inputs.

Both history stacks are initialized as zero matrices of the appropriate size. As state estimates become available, they are added, along with the corresponding control input, to $H_2$, at a predetermined time interval until $H_2$ is full. After  $H_2$ is full, any newly available state estimates are selected to replace existing state estimates in $H_2$ if the condition number of $\hat{\Sigma}^{\top}\hat{\Sigma}$, calculated using the post-replacement history stack, is smaller than the condition number of $\hat{\Sigma}^{\top}\hat{\Sigma}$ before the replacement. Once the data in $H_2$ are such that the condition number of $\hat{\Sigma}^{\top}\hat{\Sigma}$ is lower than a user-selected threshold, and a predetermined amount of time has passed since the last update of $H_1$, we set $H_1 = H_2$ and purge $H_2$ by setting it back to a zero matrix. Due to the purging algorithm, the time instances $t_i$ corresponding to the data stored in the history stack $H_1$ are piecewise constant functions of time.

The IRL method developed in this paper requires that the behavior of the expert is optimal, which implies that $u(t) = K_{Ep}x(t)$ for all $t$. Since the true values of the state are not accessible, $K_{Ep}\hat x(t_i(t)) - u(t_i(t))$ cannot be expected to be equal to $0$ for the data points stored in the history stack $H_1$. This discrepancy between $K_{Ep}\hat x(t_i(t))$ and $u(t_i(t))$  results in inaccurate estimates of equivalent solutions. Since the state estimates converge to the true state exponentially, the purging process described above ensures that the discrepancy $\max_{i=1,\cdots,N}\left\Vert K_{Ep}\hat x(t_i(t)) - u(t_i(t))\right\Vert$ is bounded by an exponentially decaying envelope, and so is the resulting inaccuracy in the estimation of an equivalent solution.


        
\section{Regularized History Stack Observer for IRL Problems with Multiple Solutions\label{sec:RHSO}}
Due to purging and improved state estimates, $\hat \Sigma$ being full rank implies that $\Sigma$ is eventually full rank, and as a result, $\Sigma W = \Sigma_u$ has a unique solution. As such, the explicit assumption that $\hat{\Sigma}$ is full rank implies an implicit assumption that the IRL problem admits a unique solution. Lack of uniqueness thus necessitates algorithms that can incorporate a rank-deficient $\hat{\Sigma}$. To that end, a regularized HSO (RHSO) is developed in this paper where the term $K_4(\hat{\Sigma}^\top\hat{\Sigma})^{-1}$ is replaced by a generic positive definite matrix
to yield 
\begin{equation}\label{Equation: K RHSO}
    K \coloneqq
    \begin{bmatrix}
        K_{3} & 0_{n\times N+Nm}\\
        0_{P_S+P_Q+M-1\times L} & K_4\hat{\Sigma}^{\top}
    \end{bmatrix},
\end{equation}
where $K_4$ is a positive definite matrix of dimension $P_S + P_Q + M - 1$.
In the following lemmas and theorems, it is shown that under a novel informativity condition on the recorded data, the modification above leads to convergence to an equivalent solution when the IRL problem admits multiple solutions and convergence to the true cost functional of the expert, up to a scaling factor, when the IRL problem admits a unique solution. While the modification itself is relatively minor, the above somewhat surprising results are the key contributions of this work. The analysis requires a data informativity condition summarized in Definition \ref{Definition: FI} below. 
\begin{definition}\label{Definition: FI}
    The signal $(\hat{x},u)$ is called \emph{finitely informative (FI)} if there exists a time instance $T >0 $ such that for some $\{t_1, t_2, \ldots , t_N\} \subset [0,T] $,
    \begin{gather}
        \spanop\left\{\hat x(t_i)\right\}_{i=1}^N = \mathbb{R}^n, \quad  \Sigma_u \in \rangeop(\hat{\Sigma}), \quad\text{and}\nonumber\\
        \spanop\left\{\hat x(t_i) \hat{x}^{\top}(t_i)\right\}_{i=1}^N = \{\mathbb{Z}\in \mathbb{R}^{n\times n}| \mathbb{Z}=\mathbb{Z}^{\top}\}.\label{Equation: Sigma_u Condition}
    \end{gather}
In addition, for a given $\epsilon > 0$, if $\min\{\eigop(X X^\top)\} > \epsilon$ and $\min\{\eigop(Z Z^\top)\} > \epsilon$, where $X \coloneqq [\hat{x}(t_1),\ldots,\hat{x}(t_N)]$, $Z \coloneqq [\uvecop(\hat x(t_1) \hat{x}^{\top}(t_1)),\ldots,\uvecop(\hat x(t_N) \hat{x}^{\top}(t_N))] \in\mathbb{R}^{\frac{n(n+1)}{2}\times N}$, and $\uvecop(\hat x(t_i) \hat{x}^{\top}(t_i)) \in \mathbb{R}^{\frac{n(n+1)}{2}}$ denotes vectorization of the upper triangular elements of the symmetric matrix $\hat x(t_i) \hat{x}^{\top}(t_i) \in\mathbb{R}^{n\times n}$, then $(\hat{x},u)$ is called \emph{$\epsilon-$finitely informative ($\epsilon-$FI)}.
\end{definition}

\begin{remark}\label{Remark: excitation}
    The three FI conditions in Definition \ref{Definition: FI} are utilized in the subsequent analysis to show that as the equivalence metric converges to zero, the corresponding weight estimates converge to an equivalent solution.
    \begin{enumerate}
        \item The condition $\spanop\left\{\hat x(t_i)\right\}_{i=1}^N = \mathbb{R}^n$ is an excitation-like condition that requires the state estimates stored in the history stack to be linearly independent. This condition is not restrictive in general, however it can fail if the system has trajectories that are confined to a subspace of dimension less than $n$. This condition can be monitored online by ensuring that the minimum eigenvalue of $X X^\top$ is strictly positive, and as shown in Fig. \ref{fig:span_cond}, it is met in the simulation study.
        \item The condition $\spanop\left\{\hat x(t_i) \hat{x}^{\top}(t_i)\right\}_{i=1}^N = \{\mathbb{Z}\in \mathbb{R}^{n\times n}| \mathbb{Z}=\mathbb{Z}^{\top}\}$ is a sufficient condition for $\hat{x}^{\top}(t_i)M \hat x(t_i)=0, \forall i=1,\cdots,N$ to imply $M=0$. It is not clear how restrictive this condition is, but it can be verified online by ensuring that the minimum eigenvalue of the matrix $ZZ^\top$ defined above is strictly positive. As shown in Fig. \ref{fig:symmetric_cond} this condition is met in the simulation study.
        \item The condition $\Sigma_u \in \rangeop(\hat{\Sigma})$ is met provided at least one set of weights $\hat W$ satisfies $\Sigma_u=\hat{\Sigma}\hat{W}$, and as such, is not restrictive. If the IRL problem has a unique solution, then this condition is trivially met whenever $N \ge P_S + P_Q + M - 1$ and $\hat{\Sigma}$ is full rank. Furthermore, this condition can be verified online using the fact that $\Sigma_u \in \rangeop(\hat{\Sigma}) \iff \rankop\left(\begin{bmatrix} \Sigma_u & \hat{\Sigma} \end{bmatrix}\right) = \rankop(\hat{\Sigma})$. Since the expert is assumed to be optimal, $\Sigma_u = \Sigma W^*$, and as a result, $\Sigma_u \in \rangeop(\Sigma)$. Due to improving state estimates and the purging algorithm, $\hat{\Sigma}$ converges to $\Sigma$, and as a result, there exists $T>0$ such that $\Sigma_u \in \rangeop(\hat\Sigma)$ for all $t\geq T$. As shown in Fig. \ref{fig:Sigma_u_cond} this condition is met in the simulation study.
    \end{enumerate}
    If the optimal trajectories of the expert do not meet the excitation conditions, an excitation signal can be added to the control input of the expert. As long as the excitation signal is known to the learner, the learner can infer the optimal control input of the expert needed to implement the developed RHSO.
\end{remark}
\begin{remark}
   In the case of noisy measurements, the feedback gains $K_3$ and $K_4\hat{\Sigma}^\top$ in \eqref{Equation: K RHSO} can be replaced by Kalman gains. While empirical evidence suggests that the use of the Kalman gain results in improved performance (see \cite[Section 2.3.3]{SCC.Town2023}), the stability guarantees in this paper are for deterministic systems with $K$ selected according to \eqref{Equation: K HSO}. Extension of the developed stability guarantees to the case where the measurements are noisy and $K$ is the Kalman gain is out of the scope of this paper.
\end{remark}
The following technical lemma is needed to prove convergence of the equivalence metric to zero.
\begin{lemma} \label{Lemma_Omega_D}
    If $\hat{\Sigma}$ and $\Sigma_u$ satisfy \eqref{Equation: Sigma_u Condition}, then $\Omega_\Delta \cap \nullop(\hat\Sigma^{\top}) = \{0\}$, where $\Omega_\Delta \coloneqq  \big\{ \Delta \in \mathbb{R}^{N(m+1)} \mid  \Delta = \Sigma_u - \hat{\Sigma} \hat{W}$, for some $\hat{W} \in \mathbb{R}^{P_S+P_Q+M-1} \big\}$.
\end{lemma}
\begin{proof}
    If $\Delta \in \nullop({\hat{\Sigma}^{\top}})$, then $\hat{\Sigma}^{\top}\Delta = 0$. In addition, if $\Delta \in \Omega_{\Delta}$, then exists a $\hat{W}$ such that $\hat{\Sigma}^{\top}\left(\Sigma_u - \hat{\Sigma}\hat{W}\right) = 0$. The FI condition in \eqref{Equation: Sigma_u Condition} implies the existence of some $W^{\prime}$ such that $\Sigma_u = \hat{\Sigma}W^{\prime} $. Therefore, $\hat{\Sigma}^{\top}\left(\hat{\Sigma}W^{\prime} - \hat{\Sigma}\hat{W}\right) = 0$. As a result, $\hat{\Sigma}W^{\prime} - \hat{\Sigma}\hat{W} \in \nullop(\hat{\Sigma}^{\top})$. By definition of the range space, $\hat{\Sigma}W^{\prime} - \hat{\Sigma}\hat{W} \in \rangeop(\hat{\Sigma})$. Since $ \rangeop(\hat{\Sigma}) = (\nullop(\hat{\Sigma}^{\top}))^{\perp} $ \cite[Section 4.1]{SCC.Strang2009}, $\hat{\Sigma}W^{\prime} - \hat{\Sigma}\hat{W} \in (\nullop(\hat{\Sigma}^{\top}))^{\perp}\cap\nullop(\hat{\Sigma}^{\top})$. Therefore, $\hat{\Sigma}W^{\prime} - \hat{\Sigma}\hat{W} = 0$, which implies that $\Delta = 0$.
\end{proof}
Theorem \ref{Theorem: Delta Convergence} below shows that for given \emph{fixed} matrices $\hat{\Sigma}$ and $\Sigma_u$ that satisfy \eqref{Equation: Sigma_u Condition}, if the weights $\hat{W}$ are updated using the update law in \eqref{Hopefully_is_update_law}, then the equivalence metric $\Delta$ converges to the origin.
\begin{theorem}\label{Theorem: Delta Convergence}
    Let $\Delta \coloneqq \Sigma_u - \Sigma \hat{W}$. If $\Sigma_u \in \nullop(\hat{\Sigma}^{\top})^\perp$, the gain $K$ is selected according to \eqref{Equation: K RHSO}, and the weights $\hat{W}$ are updated using the update law in \eqref{Hopefully_is_update_law}, then $\lim_{t\to\infty}\Delta(t) = 0$. In addition if full state information is available (i.e., $\hat{x} = x$ and as a result, $\hat\Sigma = \Sigma$),  $ \Delta = 0 $, $\spanop\{x(t_i)\}_{i=1}^{N} = \mathbb{R}^n$, $\spanop\{x(t_i) x(t_i)^{\top}\}_{i=1}^N = \{\mathbb{Z}\in \mathbb{R}^{n\times n}| \mathbb{Z}=\mathbb{Z}^{\top}\}$,
    and if the matrix $\hat{R}$, extracted from $\hat{W}$, is invertible, then the matrices $\hat{Q}$, $\hat{S}$, and $\hat{R}$, extracted from $\hat{W}$, constitute an equivalent solution of the IRL problem.
\end{theorem}
\begin{proof}
    Using the update law in \eqref{Hopefully_is_update_law}, the time-derivative of $\Delta$ can be expressed as
    \begin{equation}\label{Equation: Delta_dot}
        \dot{\Delta} = -\hat{\Sigma}K_4\hat{\Sigma}^{\top}\Delta.
    \end{equation}       
    Consider the positive definite and radially unbounded candidate Lyapunov function $V:\mathbb{R}^{N(m+1)}\to\mathbb{R}$ defined as
    \begin{equation}
        \begin{aligned}
            V(\Delta) = \frac{1}{2}\Delta^{\top}\Delta.
        \end{aligned}
    \end{equation}
    The orbital derivative of $V$ along the solutions of \eqref{Equation: Delta_dot} is given by
    \begin{equation}\label{eq:V_dot_Delta}
            \dot{V}(\Delta) = -\Delta^{\top}\hat{\Sigma}K_4\hat{\Sigma}^{\top}\Delta.
    \end{equation}
    Note that all points in null space of $\Sigma^\top$ are equilibrium points of \eqref{Equation: Delta_dot}. Since $\Sigma^\top$ is not assumed to be full rank, $\nullop(\Sigma^\top) \neq \{0\}$. As a result, if $\Sigma^\top$ is not full rank, then the origin cannot be an asymptotically stable equilibrium point of \eqref{Equation: Delta_dot}. The analysis thus requires the invariance principle.
    
    Since $\Omega_\Delta = \{\Sigma_u\} \ominus \rangeop(\hat{\Sigma})$, where $\ominus$ denotes the Minkowski difference, it is easy to see that provided \eqref{Equation: Sigma_u Condition} holds, $\Omega_\Delta$ is a subspace of $\mathbb{R}^{N(m+1)}$. Indeed, given $\alpha,\beta\in\mathbb{R}$ and $\Delta_1,\Delta_2\in\Omega_\Delta$, with $\Delta_i = \Sigma_u - \hat{\Sigma}\hat{W}_i$ for $i=1,2$, we have $\alpha\Delta_1 + \beta\Delta_2 = \hat\Sigma_u + (\alpha + \beta - 1)\Sigma_u - \hat{\Sigma}(\alpha\hat{W}_1+\beta\hat{W}_2)$. If \eqref{Equation: Sigma_u Condition} holds, then $\Sigma_u\in\rangeop{\hat\Sigma}$, and as a result, $\alpha\Delta_1 + \beta\Delta_2\in\Omega_\Delta$. Since $\Omega_\Delta$ is a subspace of a finite dimensional topological space, it is also closed.
    
    If $\Delta_0 \in \Omega_\Delta$ then there exists $\hat{W}_0$ such that $\Delta_0 = \Sigma_u - \hat{\Sigma} \hat{W}_0$. Let $t\mapsto \hat{W}_{\Delta_0}(t)$ be a solution of \eqref{Hopefully_is_update_law} starting from $\hat{W}_0$ with the interval of existence $\mathcal{I}$. For almost all $t \in \mathcal{I}$, we have $\dot{\hat{W}}_{\Delta_0} = K_4\hat{\Sigma}^\top(\Sigma_u - \hat{\Sigma} \hat{W}_{\Delta_0})$, which implies $\frac{\mathrm{d}}{\mathrm{d}t}(\Sigma_u - \hat{W}_{\Delta_0}) = -K_4\hat{\Sigma}^\top(\Sigma_u - \hat{\Sigma} \hat{W}_{\Delta_0})$. Letting $\Delta_{\Delta_0} = \Sigma_u - \hat{\Sigma}\hat{W}_{\Delta_0}$, it can be concluded that for almost all $t\in \mathcal{I}$, $\dot{\Delta}_{\Delta_0}(t) = -K_4\hat{\Sigma}^\top\Delta_{\Delta_0}(t)$. That is, $t\mapsto \Delta_{\Delta_0}(t)$ is a solution of \eqref{Equation: Delta_dot} on $\in \mathcal{I}$, starting from $\Delta_0$. Uniqueness of solutions then implies that $t\mapsto \Delta_{\Delta_0}(t)$ is the only solution of \eqref{Equation: Delta_dot} on $\in \mathcal{I}$ starting from $\Delta_0$. Using continuity of $t\mapsto \Delta_{\Delta_0}(t)$ along with the facts that $\Omega_\Delta$ is closed and $\Delta_{\Delta_0}(t) \in \Omega_\Delta$ for almost all $t\in \mathcal{I}$, it can be concluded that $\Delta_{\Delta_0}(t) \in \Omega_\Delta$ for all $t\in \mathcal{I}$. As a result, $\Omega_\Delta$ is positively invariant with respect to \eqref{Equation: Delta_dot}.
    
    For any $c>0$, the sublevel set $\Omega_c:=\{\Delta \in \mathbb{R}^{N(m+1)}| V(\Delta) \le c\}$ is compact. From \eqref{eq:V_dot_Delta}, we conclude that $\Omega_c$ is positively invariant with respect to \eqref{Equation: Delta_dot}. 
    As a result, $\Omega := \Omega_c \cap \Omega_\Delta$ is also positively invariant with respect to \eqref{Equation: Delta_dot}. Since $\Omega_c$ is compact and $\Omega_\Delta$ is closed, $\Omega$ is also compact. The invariance principle \cite[Theorem 4.4]{SCC.Khalil2002} can thus be invoked to conclude that all trajectories starting in $\Omega$ converge to the largest invariant subset of $\{ \Delta \in \Omega \mid \dot{V}(\Delta)=0\}$.
    
    
    The set $\{\Delta \in \Omega | \dot{V}(\Delta) = 0\}$, is equal to $\nullop(\hat{\Sigma}^{\top}) \cap \Omega$ as $\hat{\Sigma}^{\top}\Delta=0$ only when $\Delta \in \nullop(\hat{\Sigma}^{\top})$.
    Furthermore, from Lemma \ref{Lemma_Omega_D}, provided $\Sigma_u \in (\nullop(\hat\Sigma^{\top}))^{\perp}$, the only $\Delta$ that can be a member of $\nullop(\hat{\Sigma}^{\top}) \cap \Omega_{\Delta}$ is $\Delta = 0$.
    Since the set $\{0\}$ is positively invariant with respect to \eqref{Equation: Delta_dot}, it is also the largest invariant subset of $\{\Delta \in \Omega | \dot{V}(\Delta) = 0\}$.
    
    As a result, by the invariance principle, all trajectories that start in $\Omega$
    converge to the origin.
    Since V is radially unbounded, $\Omega_c$ can be selected to be large enough to include any initial condition in $\Omega_\Delta$. Thus, all solutions of \eqref{Equation: Delta_dot} that start in $\Omega_\Delta$
    converge to the origin. In particular, $\Delta$ converges to zero along the solutions of the update law in \eqref{Hopefully_is_update_law}.        
         
    To prove equivalence when $\Delta = 0$, the equality $\hat{R}^{-1}B^{\top} \hat{S} = K_{Ep}$ must be established. Indeed, if $\{x(t_i)\}_{i=1}^{N}$ spans $\mathbb{R}^n$ there is a unique matrix $K$ that satisfies $u(t_i) = K x(t_i)$ for all $i = 1, \ldots , N$. Letting $U = [u(t_1) , \ldots , u(t_N)]$ and $X = [x(t_1) , \ldots , x(t_N)]$, this unique matrix is given by $K = U X^{\top}(X X^{\top})^{-1}$. It is also known that because the behavior of the expert is optimal, the observed data satisfy $u(t_i) = -K_{Ep} x(t_i)$ for all $i = 1 , \ldots , N$. Since $\Delta = 0$, the observed data points satisfy $u(t_i) = -\hat{R}^{-1}B^{\top} \hat{S} x(t_i)$ for all $i = 1 , \ldots , N$. Since there is only one matrix $K$ that satisfies $u(t_i) = -K x(t_i)$ for all $i = 1,\ldots,N$, all three of the matrices above must be equal, i.e., $K = K_{Ep} = \hat{R}^{-1}B^{\top} \hat{S}$.

    The fact that if $\Delta=0$ then $x(t_i)^{\top}\left(A^{\top}\hat S + \hat SA-\hat SB\hat R^{-1}B^{\top}\hat S+\hat Q\right)x(t_i)=0$ holds for all points in $H_1$ is immediate from the construction of $\Delta$.
    Furthermore, with a slight modification of the proof from \cite{SCC.StackExchange.Quadratic2019}, ($\hat{Q}$, $\hat{S}$, $\hat{R}$) can be proven to satisfy the ARE if $\Delta = 0$ and $\{x(t_i) x(t_i)^{\top}\}_{i=1}^N$ spans all symmetric matrices. To that end, let $e_i$ be the basis vector of zeros with a one in the $i^{th}$ position such that $e_je_k^{\top} + e_ke_j^{\top}=\sum_{i=1}^{N}\alpha_ix(t_i)x(t_i)^{\top}$ for some $\alpha_1 \cdots \alpha_N \in \mathbb{R}$. Rewriting \eqref{HJB} with $\hat{M}=\left(A^{\top}\hat S + \hat SA-\hat SB\hat R^{-1}B^{\top}\hat S+\hat Q\right)$, $\sum_{i=1}^{N}\alpha_ix(t_i)^{\top}\hat{M}x(t_i) = \sum_{i=1}^{N}\sum_{p=1}^{n}\sum_{q=1}^{n}\alpha_ix_{i,p}\hat{M}_{p,q}x_{i,q} = \sum_{i=1}^{N}\sum_{p=1}^{n}\hat{M}_{p,q}\sum_{q=1}^{n}\alpha_ix_{i,p}x_{i,q} $.
    Now, for any fixed $j,k$, select $\{\alpha_i\}_{i=1}^{N}$ such that $\sum_{i=1}^{N}\alpha_ix(t_i)x(t_i)^{\top}=e_je_k^{\top}+e_ke_j^{\top}$, where 
    \[
        \sum_{i=1}^{N}\alpha_ix(t_i)x(t_i)^{\top}=
        \begin{cases}
            1 & \text{if } p=j, q=k, \\
            1 & \text{if } p=k, q=j, \\
            0 & \text{otherwise}.
        \end{cases}
    \]
    As a result, $\sum_{i=1}^{N}\sum_{p=1}^{n}\hat{M}_{p,q}\sum_{q=1}^{n}\alpha_ix_{i,p}x_{i,q}=e_k^{\top}\hat{M}e_j+e_j^{\top}\hat{M}e_k=\hat{M}_{j,k}+\hat{M}_{k,j}=2\hat{M}_{j,k}=0.$
    Since $j$ and $k$ were arbitrary, $\hat{M} = 0$. That is, the tuple ($\hat{Q}$, $\hat{S}$, $\hat{R}$) satisfies the ARE and constitutes an equivalent solution of the IRL problem. 
\end{proof}
\begin{remark}\label{Remark: projection}
    The invertibility of $\hat{R}$ is needed for $\hat K_P$ to be well-defined. While this is difficult to ensure a priori in general, it can be guaranteed in the specific case where $R$ is diagonal by using a projection operator to ensure that all diagonal elements of $\hat{R}$ remain positive. In this case, the weights are updated using the update law $\dot{\hat{W}} = \projop\left(K_4\hat{\Sigma}^\top\Delta\right)$, where $\projop(\cdot)$ denotes smooth projection (see Appendix E of \cite{SCC.Krstic.Kanellakopoulos.ea1995}) onto the convex set $\mathbb{R}^{P_S} \times \mathbb{R}^{P_Q} \times \mathbb{R}^{m-1}_{\geq \kappa}$, where  $\mathbb{R}^{m-1}_{\geq \kappa}$ denotes the set of $(m-1)-$dimensional vectors that are element-wise larger than $\kappa$ and $\kappa>0$ is a lower bound for the diagonal entries of $R$. The resulting Lyapunov derivative is $\dot{V}(\Delta) = -\Delta^\top\hat{\Sigma}\projop\left(K_4\hat{\Sigma}^\top\Delta\right)$. Invoking Lemma E.1 from \cite{SCC.Krstic.Kanellakopoulos.ea1995}, it can be concluded that $\dot{V}(\Delta) \leq -\Delta^\top\hat{\Sigma}K_4\hat{\Sigma}^\top\Delta$. The rest of the analysis then remains unchanged.
\end{remark}
Theorem \ref{Theorem: Delta Convergence} can be used to obtain the final result summarized in the definition and the theorem below.  
\begin{definition}\label{Definition: Approximate Equivalent Solution}
    Given $\varpi \geq 0$ A solution ($\hat{Q}$, $\hat{S}$, $\hat{R}$) to the IRL problem is called an $\varpi-$equivalent solution of the IRL problem if $\left\Vert \hat{M}\right\Vert\leq \varpi$, where $\hat{M}=A^{\top}\hat S+\hat S A- \hat SB\hat R^{-1}B^{\top}\hat S+\hat Q$, and optimization of the performance index $J$, with $Q=\hat{Q}$ and $R=\hat{R}$, results in a feedback matrix, $\hat{K}_p \coloneqq \hat{R}^{-1}B^{\top} \hat{S}$, that satisfies $\left\Vert \hat{K}_p - K_{Ep}\right\Vert \leq \varpi$.
\end{definition}
Due to the purging algorithm described in Section \ref{subsec:HSO}, the time instances $t_i$ corresponding to the data stored in the history stack $H_1$ are piecewise constant functions of time, where $t_1(t)$ denotes the time instance when the oldest datum in the history stack was recorded. The corollary below requires $\mathrm{lim}\,\mathrm{inf}_{t\to\infty} t_1(t)$ to be large enough, which translates into the requirement that the excitation in the trajectories of the expert lasts long enough to allow sufficiently many purging events. 

The exact lower bound on $\mathrm{lim}\,\mathrm{inf}_{t\to\infty} t_1(t)$ needed for convergence to a $\varpi-$equivalent solution is characterized in the proof of Theorem \ref{Theorem: Approximate Equivalence} below. The lower bound depends on the value of $\varpi$, the norm of the feedback gain $K_{Ep}$ of the expert, the user-selected poles of $A-K_3C$, the user-selected gain matrix $K_4$, the condition numbers of the data matrices $X$ and $Z$ introduced in Definition \ref{Definition: FI}. If $(\hat{x},u)$ is $\epsilon$-FI, the lower bounds $\min\{\eigop(X(t) X(t)^\top)\} > \epsilon$ and $\min\{\eigop(Z(t) Z(t)^\top)\} > \epsilon$, for some $\epsilon > 0$ and all $t\geq \underline{T}$, can be easily ensured using a modified history stack management algorithm that maximizes the minimum eigenvalues of $X(t) X(t)^\top$ and $Z(t) Z(t)^\top$.
\begin{theorem}\label{Theorem: Approximate Equivalence}
    Let $\underline{T} \geq 0$ denote the first time instant when $H_1$ is updated. Given $\varpi > 0$ if $\mathrm{lim}\,\mathrm{inf}_{t\to\infty} t_1(t)$ is large enough, $\Sigma_u (t) \in \nullop(\hat{\Sigma}^{\top}(t))^\perp$ for all $t\geq \underline{T}$, $K_3$ is selected so that $A - K_3C$ is Hurwitz, $\min\{\eigop(X(t) X(t)^\top)\} > \epsilon$ and $\min\{\eigop(Z(t) Z(t)^\top)\} > \epsilon$, for some $\epsilon > 0$ and all $t\geq \underline{T}$, with $X$ and $Z$ as introduced in Definition \ref{Definition: FI}, and if there exist a constant $0\leq \underline{R} < \infty$ such that the matrix $\hat{R}(t)$, extracted from $\hat{W}(t)$ is invertible with $\|\hat R^{-1}(t)\|\leq \underline{R}$ for all $t\geq \underline{T}$, then the matrices $\hat{Q}$, $\hat{S}$, and $\hat{R}$, extracted from $\hat{W}$, converge to a $\varpi-$equivalent solution of the IRL problem.
\end{theorem}
\begin{proof}
    The dynamics in \eqref{Equation: Delta_dot} ensure that $\Delta(t)$ is bounded for all $t$. The control residual error established in \eqref{CRE} can be manipulated into the form $\sigma_{\Delta_u^\prime}\left(\hat{x}(t_i(t)),u(t_i(t))\right)\hat{W}^\prime(t) = \hat R(t) \left(\tilde{K}_P(t)\hat{x}(t_i(t))+K_{Ep}\tilde{x}(t_i(t)) \right)$, where $\tilde{K}_P(t)\coloneqq \hat{R}^{-1}(t)B^{\top} \hat{S}(t) - K_{Ep}$ and $\tilde{x}(t_i(t)) \coloneqq x(t_i(t))-\hat{x}(t_i(t))$.
    Using the triangle inequality $\left\Vert \tilde{K}_P(t)\hat{x}(t_i(t))\right\Vert \le \left\Vert \hat{R}^{-1}(t)\sigma_{\Delta_u^\prime}\left(\hat{x}(t_i(t)),u(t_i(t))\right)\hat{W}^\prime(t)\right\Vert+\left\Vert K_{Ep}\tilde{x}(t_i(t)) \right\Vert$. 
    
    Note that if $\spanop\{\hat{x}(t_i(t))_{i=1}^N \} = \mathbb{R}^n$, and in particular, if $\min\{\eigop(X(t) X(t)^\top)\} > \epsilon$ then $\exists c > 0$, independent of $t$, such that $\left\Vert \tilde{K}_P(t)\hat{x}(t_i(t))\right \Vert \le \frac{\varpi}{c}, \forall i, $ implies $\left \Vert \tilde{K}_P(t) \right \Vert \le \varpi$. Select $\overline{T}_1$ large enough such that the equivalence metric $\Delta(t)$ satisfies $\left\Vert\sigma_{\Delta_u^\prime}\left(\hat{x}(t_i(t)),u(t_i(t))\right)\hat{W}^\prime(t)\right\Vert \leq \frac{\varpi }{2c\underline{R}}$, for all $i$ and for all $t\geq \overline{T}_1$. Such a $\overline{T}_1$ exists since by Theorem \ref{Theorem: Delta Convergence}, $\lim_{t\to\infty}\Delta(t) = 0$. Select $\overline{T}_2$ large enough so that the state estimation error $\tilde{x}(t_i(t))$ satisfies $\left\Vert\tilde x(t_i(t))\right\Vert \leq \frac{\varpi}{2c\left\Vert K_{Ep} \right\Vert}$ for all $t\geq \overline{T}_2$. Since $\lim_{t\to\infty}\tilde{x}(t) = 0$, existence of of such a $\overline{T}_2$ follows if $t_1(\overline{T}_2)$ is large enough. Letting $\overline{T} = \max \{\overline{T_1}, \overline{T}_2\}$, it can be concluded that for all $t\geq \overline{T}$, $\left\Vert \tilde{K}_P(t)\hat{x}(t_i(t))\right \Vert \le \frac{\varpi}{c}$, which implies $\left \Vert \tilde{K}_P(t) \right \Vert \le \varpi$.
    
    The inverse Bellman error established in \eqref{IBE} can be manipulated into $\sigma_{\delta^\prime}\left(\hat{x}(t_i(t)),u(t_i(t))\right)\hat{W}^\prime(t) = \hat{x}^{\top}(t_i(t))\hat M\hat{x}(t_i(t)) + g\left(\hat{K}_P(t), \hat{x}(t_i(t)), K_{Ep}, x(t_i(t))\right)$, where the function $g$ satisfies\footnote{For a positive function $g$, $f=O(g)$ if there exists a constant $M$ such that $\left \Vert f(x) \right\Vert \le Mg(x), \forall x$} $g=O\left(\left\Vert\tilde{K}_{P}(t)\right\Vert + \left\Vert \tilde{x}(t_i(t)) \right \Vert \right)$. Using the triangle inequality, $\left\vert \hat{x}^{\top}(t_i(t))\hat M(t)\hat{x}(t_i(t)) \right\vert \leq \left\vert\sigma_{\delta^\prime}\left(\hat{x}(t_i(t)),u(t_i(t))\right)\hat{W}^\prime\right\vert + \left\vert g\left(\hat{K}_P(t), \hat{x}(t_i(t)), K_{Ep}, x(t_i(t))\right)\right\vert$, where $\hat{M}(t)=A^{\top}\hat S(t) + \hat S(t)A-\hat S(t)B\hat R^{-1}(t)B^{\top}\hat S(t)+\hat Q(t)$
    
    Since $g=O\left(\left\Vert\tilde{K}_{P}(t)\right\Vert + \left\Vert \tilde{x}(t_i(t)) \right \Vert \right)$ and $\left\vert\sigma_{\delta^\prime}\left(\hat{x}(t_i(t)),u(t_i(t))\right)\hat{W}^\prime\right\vert \leq\left\Vert\Delta(t)\right\Vert$, a construction similar to the one in the previous paragraph can be used to show that given any $\varepsilon > 0$, that there exists a $\overline{T}$ such that for all $t \geq \overline{T}$ and for all $i = 1,\ldots, N$, $\left\vert \hat{x}^{\top}(t_i(t))\hat M(t)\hat{x}(t_i(t)) \right\vert \leq \varepsilon$.
   
    Equivalence of matrix norms implies that there exists $c>0$, independent of $t$, such that if $\left \vert \hat{M}_{j,k}(t)\right \vert\le \varpi/c $ for all $j,k = 1,\cdots,n$, then $\left\Vert \hat{M}(t) \right\Vert \le \varpi$. As a result, to complete the proof of the theorem, it suffices to construct a $\overline{T}$ such that for all $t\geq \overline{T}$ and for all $j,k  = 1,\cdots,n$, $\left \vert \hat{M}_{j,k}(t)\right \vert\le \frac{\varpi}{c}$. To construct such a $\overline{T}$, an $\varepsilon$ is constructed such that $\left\vert \hat{x}^{\top}(t_i(t))\hat M(t)\hat{x}(t_i(t)) \right\vert \leq \varepsilon, i = 1,\ldots, N$ implies $\left \vert \hat{M}_{j,k}(t)\right \vert\le \frac{\varpi}{c}, \forall j,k  = 1,\cdots,n$. Existence of the required $\overline{T}$ then follows from the discussion in the previous paragraph.
    
    Let $e_i$ be the basis vector of zeros with a one in the $i^{th}$ position. For a fixed $j$ and $k$, selecting constants $\alpha_{1,j,k} \cdots \alpha_{N,j,k} \in \mathbb{R}$ and rewriting \eqref{HJB}, we have \begin{multline*}
        \sum_{i=1}^{N}\alpha_{i,j,k}\hat x^{\top}(t_i(t))\hat{M}(t)x(t_i(t)) = \sum_{i=1}^{N}\sum_{p=1}^{n}\sum_{q=1}^{n}\alpha_{i,j,k}\hat x_{p}(t_i(t))\hat{M}_{p,q}(t)\hat x_{q}(t_i(t)) = \\\sum_{i=1}^{N}\sum_{p=1}^{n}\hat{M}_{p,q}(t)\sum_{q=1}^{n}\alpha_{i,j,k}\hat x_{p}(t_i(t))\hat x_{q}(t_i(t)).
    \end{multline*} If $\spanop\{\hat{x}(t_i(t)) \hat{x}^{\top}(t_i(t))\}_{i=1}^N = \{\mathbb{Z}\in \mathbb{R}^{n\times n}| \mathbb{Z}=\mathbb{Z}^{\top}\}$, then for any fixed $j,k$, we can select $\{\alpha_{i,j,k}(t)\}_{i=1}^{N}$ such that $\sum_{i=1}^{N}\alpha_{i,j,k}(t)\hat{x}(t_i(t))\hat{x}^{\top}(t_i(t))=e_je_k^{\top}+e_ke_j^{\top}$, that is, the $(p,q)$ element of $\sum_{i=1}^{N}\alpha_{i,j,k}(t)\hat{x}(t_i(t))\hat{x}^{\top}(t_i(t))$ is 1 if $p=j$ and $q=k$, it is also 1 if $p=k$ and $q=j$, and it is zero otherwise. As a result, $\sum_{i=1}^{N}\sum_{p=1}^{n}\hat{M}_{p,q}(t)\sum_{q=1}^{n}\alpha_{i,j,k}(t)\hat x_{p}(t_i(t))\hat x_{q}(t_i(t)) = e_k^{\top}\hat{M}(t)e_j+e_j^{\top}\hat{M}(t)e_k=\hat{M}_{j,k}(t)+\hat{M}_{k,j}=2\hat{M}_{j,k}(t)$.
    If $\min\{\eigop(Z(t) Z(t)^\top)\} > \epsilon$ then the coefficients $\alpha_{i,j,k}$ are bounded such that $\sup_{t\geq \underline{T}}\max_{i,j,k}(\{|\alpha_{i,j,k}(t)|\}_{i,j,k=1}^{N,n,n}) \leq\alpha < \infty$ for some $\alpha > 0$. 
    
    Select $\varepsilon = \frac{2\varpi}{c\alpha N}$ and note that $ \left\Vert \hat x^{\top}(t_i(t))\hat{M}(t)\hat x(t_i(t))\right\Vert \leq \frac{2\varpi}{c\alpha N},\forall i=1,\cdots,N$ implies that for all $j,k=1,\ldots,n$,
    \begin{multline*}
        \left\vert2\hat{M}_{j,k}(t)\right\vert = \left\vert \sum_{i=1}^{N}\alpha_{i,j,k}(t) \hat{x}^{\top}(t_i(t))\hat{M}(t)\hat{x}(t_i(t)) \right\vert \leq\\
        \alpha N \max_i\left(\left\{\left\Vert \hat{x}^{\top}(t_i(t))\hat{M}(t)\hat{x}(t_i(t))\right\Vert\right\}_{i=1}^N\right) \leq \frac{2\varpi}{c},
    \end{multline*}
    which implies that for all $j,k=1,\ldots,n$, $\left\vert \hat M_{j,k}(t)\right\vert \leq \frac{\varpi}{c}$, which completes the proof of the theorem. 
\end{proof}
\section{Simulations}\label{Section: Simulations}
\subsection{Methods and Results}\label{Section: Non-Unique}
To demonstrate the ability of the developed method to obtain equivalent solutions to IRL problems that admit multiple solutions, an IRL problem that has a product structure is constructed and linearly transformed. The results in \cite{SCC.Jean.Maslovskaya2018} ensure that the resulting transformed IRL problem admits multiple solutions.

The state space model is given by
\begin{gather*}
    A = \begin{bmatrix}
            -0.2 & 0.4 & 1.6\\
            3.7 & 1.6 & -3.1\\
            -3.2 & 0.4 & 4.6
        \end{bmatrix}, B = \begin{bmatrix}
            1 & 2 & -1\\
            -1 & 3 & 4\\
            1 & 2 & -3
        \end{bmatrix}, C = \begin{bmatrix}
            1.7 & -0.4 & -1.1\\
            -0.1 & 0.2 & 0.3\\
            0.5 & 0 & -0.5
        \end{bmatrix}.
\end{gather*}
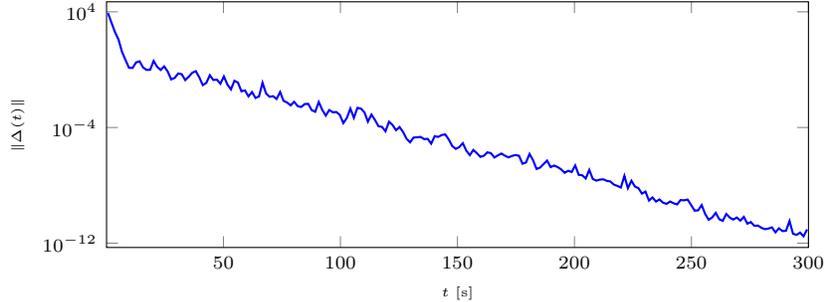
\begin{figure}[t]
    \centering
    \begin{tikzpicture}
    \begin{semilogyaxis}[
        xlabel={$t$ [s]},
        ylabel={$\left\Vert\Delta(t)\right\Vert$},
        label style={font=\tiny},
        tick label style={font=\scriptsize},
        legend pos = north east,
        legend style={nodes={scale=0.75, transform shape}},
        enlarge y limits=0.05,
        enlarge x limits=0.002,
        width=0.9\linewidth,
        height=0.4\linewidth,
    ]
        \addplot+ [thick, mark=none] table [x index=0, y index=1] {data/Delta_norm.dat};
    \end{semilogyaxis}
\end{tikzpicture}
    \caption{A log-scale plot of the 2-norm of $\Delta$ as a function of time.}
    \label{fig:Toy_Delta_non}
\end{figure}
\begin{figure}
    \centering
    \begin{tikzpicture}
    \begin{semilogyaxis}[
        xlabel={$t$ [s]},
        ylabel={$\left\Vert\hat{K}_p(t) - K_{Ep}\right\Vert$},
        label style={font=\tiny},
        tick label style={font=\scriptsize},
        legend pos = north east,
        legend style={nodes={scale=0.75, transform shape}},
        enlarge y limits=0.05,
        enlarge x limits=0,
        width=0.9\linewidth,
        height=0.4\linewidth,
    ]
        \addplot+ [thick, mark=none] table [x index=0, y index=1] {data/kp_error.dat};
    \end{semilogyaxis}
\end{tikzpicture}
    \caption{A log-scale plot of the induced 2-norm of the error between the estimated feedback gain and the feedback gain of the expert as a function of time.}
    \label{fig:Toy_Kp_non}
\end{figure}
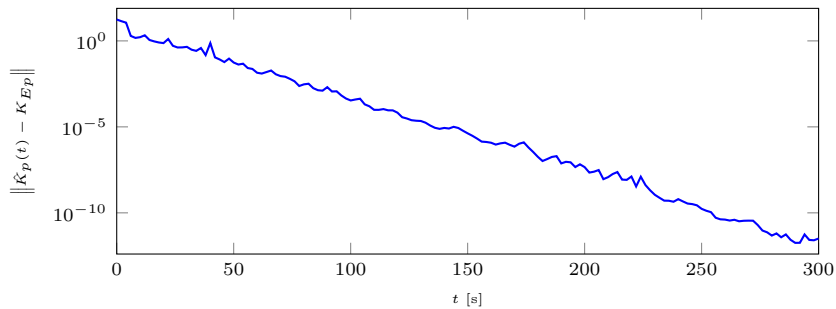
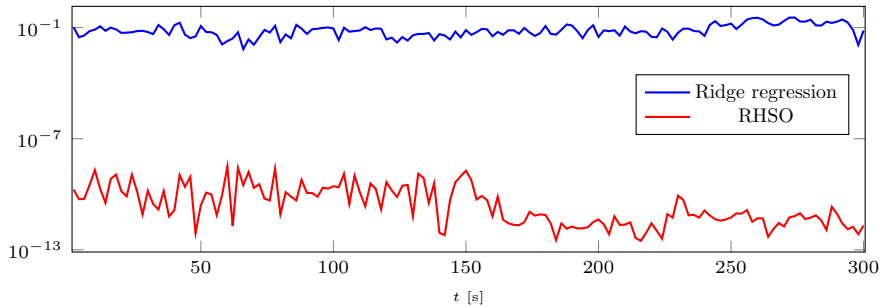
\begin{figure}[t]
    \centering
    \begin{tikzpicture}
    \begin{semilogyaxis}[
        xlabel={$t$ [s]},
        label style={font=\tiny},
        tick label style={font=\scriptsize},
        legend style={
            nodes={scale=0.75, transform shape},
            at={(0.98,0.6)},
            anchor=east
        },
        enlarge y limits=0.05,
        enlarge x limits=0.002,
        width=\linewidth,
        height=0.4\linewidth,
    ]
        \addplot+ [thick, mark=none] table [x index=0, y index=1] {data/x_norm_error_ridge.dat};
        \addplot+ [thick, mark=none] table [x index=0, y index=1] {data/x_norm_error.dat};
        \legend{Ridge regression,RHSO}
    \end{semilogyaxis}
\end{tikzpicture}
    \caption{A log-scale plot of the 2-norm of the error between the state trajectory of the expert and the state trajectory of the learner under the learned feedback gain for a problem that admits multiple solutions. The red trajectory corresponds to the feedback gain learned using the RHSO and the blue trajectory corresponds to the feedback gain computed using offline ridge regression.}
    \label{fig:ridge_comparison}
\end{figure}
The expert implements a feedback policy that minimizes the cost functional in \eqref{cost_fun} with\footnote{The notation $\diag(v)$ represents a diagonal matrix with the elements of the vector $v$ along the diagonal.}
\begin{equation}
    Q = \begin{bmatrix}
        12.32 & -2.74 & -8.26\\
        -2.74 & 0.68 & 1.82\\
        -8.26 & 1.82 & 5.68
    \end{bmatrix},\quad
    R = \begin{bmatrix}
        1 & 0 & 0\\
        0 & 4 & 0\\
        0 & 0 & 7
    \end{bmatrix}.
\end{equation}
To ensure that the history stack satisfies the sufficient condition in \eqref{Equation: Sigma_u Condition}, an excitation signal comprised of a sum of $20$ sinusoidal signals is added to the input of the expert in \eqref{Equation: Linear System}. The magnitudes are set to $0.5$ and the frequencies and phases are randomly selected from the ranges $0.001\,\mathrm{Hz}$ to $1\,\mathrm{Hz}$ and $0\,\mathrm{rad}$ to $\pi\,\mathrm{rad}$, respectively. Since the regressor $\hat\Sigma$ is a nonlinear function of $\hat x$, a precise characterization of the excitation signal needed to satisfy the finite informativity conditions in Definition \ref{Definition: FI} is difficult to obtain. Drawing inspiration from persistence of excitation results for linear regressors, the number of frequencies is selected to be higher than the number of unknown parameters, which in this example is 14. The excitation signal is assumed to be known to the learner, so it can be subtracted from the total input of the expert to infer the optimal input of the expert.
        
To facilitate comparison with ridge regression, the matrix $K_4$ is selected as $K_4 = (\hat{\Sigma}^{\top}\hat{\Sigma}+\epsilon I)^{-1}$. Data are added to the history stack every 0.05 seconds and the history stack is purged if it is full and either the condition number of $\hat{\Sigma}^{\top}\hat{\Sigma}+\epsilon I$ is smaller than $1\times10^5$, or 2 seconds have elapsed since the last purge.\footnote{See \cite{SCC.Kamalapurkar2018} for further details on condition number minimization.} The weights are $\hat{W}$ are randomly sampled from a standard normal distribution.

A Luenberger observer is utilized for state estimation by selecting the gain $K_3$ to place the poles of $(A-K_3C)$ at $p_1 = -0.1$, $p_2 = -1.5$ and $p_3 = -2$ using the MATLAB ``place'' command. These values are selected by trial and error to achieve a sufficiently fast convergence rate for the Luenberger observer. The parameters of the RHSO are held constant for all simulations in this paper unless otherwise stated.

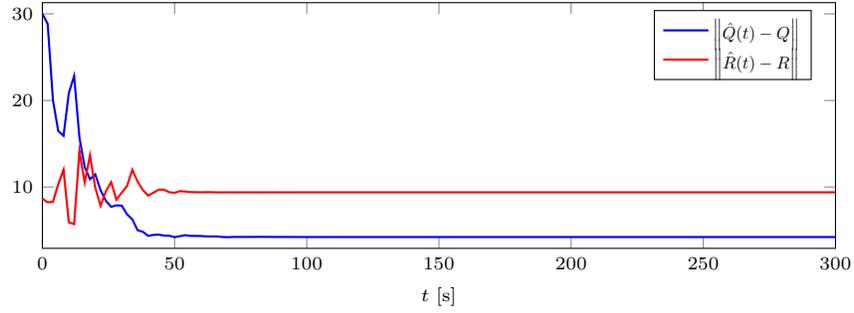
\begin{figure}
    \centering
    \begin{tikzpicture}
    \begin{axis}[
        xlabel={$t$ [s]},
        label style={font=\scriptsize},
        tick label style={font=\scriptsize},
        legend pos = north east,
        legend style={nodes={scale=0.65, transform shape}},
        enlarge y limits=0.05,
        enlarge x limits=0,
        width=\linewidth,
        height=0.4\linewidth,
    ]
        \addplot+ [thick, mark=none] table [x index=0, y index=1] {data/Q_norm_error.dat};
        \addplot+ [thick, mark=none] table [x index=0, y index=1] {data/R_norm_error.dat};
        \legend{$\left\Vert\hat{Q}(t) - Q\right\Vert$, $\left\Vert\hat{R}(t) - R\right\Vert$}
    \end{axis}
\end{tikzpicture}           
    \caption{A plot of the induced 2-norm of the error between the estimated $\hat{Q}$ (red) and $\hat{R}$ (blue) matrices and the $Q$ and $R$ matrices of the expert as a function of time.}
    \label{fig:Toy_QR_Diff_non}
\end{figure}

Fig. \ref{fig:Toy_Delta_non} demonstrates the convergence of $\Delta$ to the origin as per Theorem \ref{Theorem: Delta Convergence} and  Fig. \ref{fig:Toy_Kp_non} demonstrates the convergence of the estimated feedback gain to a neighborhood of the feedback matrix of the expert, as per Theorem \ref{Theorem: Approximate Equivalence}. Finally, Fig. \ref{fig:Toy_QR_Diff_non} indicates that the cost functional converges to a functional that is different from that of the expert, confirming that the IRL problem under consideration admits multiple equivalent solutions.

\begin{figure}
    \centering
    \begin{tikzpicture}
    \begin{axis}[
        xlabel={$t$ [s]},
        label style={font=\scriptsize},
        tick label style={font=\scriptsize},
        legend pos = north east,
        legend style={nodes={scale=0.75, transform shape}},
        enlarge y limits=0.05,
        enlarge x limits=0.002,
        ymax = 1,
        ymin = 0,
        width=\linewidth,
        height=0.4\linewidth,
    ]
        \addplot+ [only marks] table [x index=0, y index=1] {data/symmetric_span_condition.dat};
    \end{axis}
\end{tikzpicture}
    \caption{This plot is equal to 1 if $\spanop\{\hat{x}(t_i)\hat{x}^{\top}(t_i)\}_{i=1}^N = \{\mathbb{Z}\in\mathbb{R}^{n\times n}\mid \mathbb{Z} = \mathbb{Z}^{\top}\}$ and 0 otherwise.}
    \label{fig:symmetric_cond}
\end{figure}
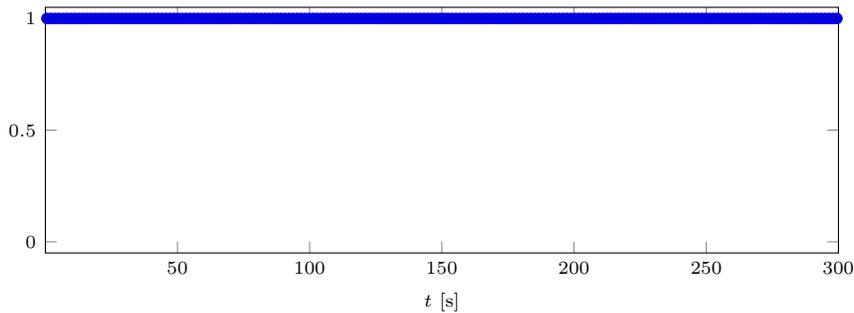

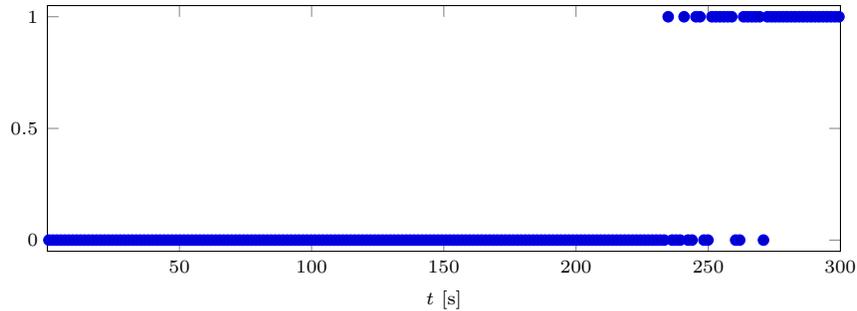
\begin{figure}
    \centering
    \begin{tikzpicture}
    \begin{axis}[
        xlabel={$t$ [s]},
        label style={font=\scriptsize},
        tick label style={font=\scriptsize},
        legend pos = north east,
        legend style={nodes={scale=0.75, transform shape}},
        enlarge y limits=0.05,
        enlarge x limits=0.002,
        width=\linewidth,
        height=0.4\linewidth,
    ]
        \addplot+ [only marks] table [x index=0, y index=1] {data/Sigma_u_condition.dat};
    \end{axis}
\end{tikzpicture}
    \caption{This plot is equal to 1 if $\Sigma_u(t)\in\rangeop(\hat{\Sigma}(t))$ and 0 otherwise.}
    \label{fig:Sigma_u_cond}
\end{figure}

Like most excitation conditions in reinforcement learning, this excitation condition cannot be guaranteed \textit{a priori}. The best practice is to monitor whether it is met online. To examine whether the sufficient conditions detailed in Definition \ref{Definition: FI} hold, stem plots are generated that equal 1 when the conditions hold and 0 when they do not (see Figs. \ref{fig:span_cond}, \ref{fig:symmetric_cond}, and \ref{fig:Sigma_u_cond}).       

\begin{figure}
    \centering
    \begin{tikzpicture}
    \begin{axis}[
        xlabel={$t$ [s]},
        label style={font=\tiny},
        tick label style={font=\scriptsize},
        legend pos = north east,
        legend style={nodes={scale=0.75, transform shape}},
        enlarge y limits=0.05,
        enlarge x limits=0.002,
        ymax = 1,
        ymin = 0,
        width=\linewidth,
        height=0.4\linewidth,
    ]
        \addplot+ [only marks] table [x index=0, y index=1] {data/x_span_condition.dat};
    \end{axis}
\end{tikzpicture}
    \caption{This plot is equal to 1 if $\spanop\{\hat{x}(t_i(t))\}_{i=1}^N = \mathbb{R}^n$ and 0 otherwise.}
    \label{fig:span_cond}
\end{figure}
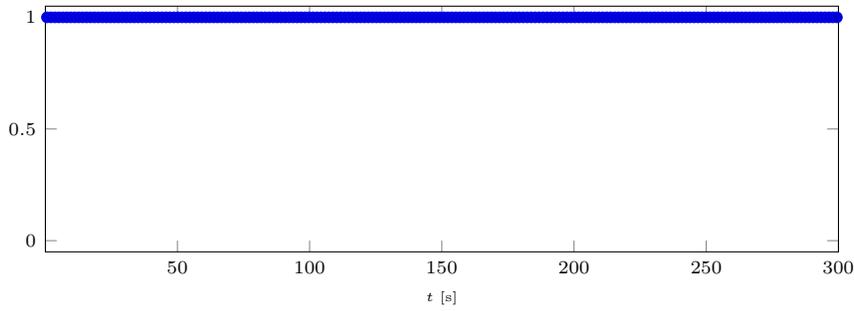

\subsection{A linear IRL problem with a unique solution}\label{Section: Unique}
\begin{figure}
    \centering
    \begin{tikzpicture}
    \begin{axis}[
        xlabel={$t$ [s]},
        label style={font=\scriptsize},
        tick label style={font=\scriptsize},
        legend pos = north east,
        legend style={nodes={scale=0.75, transform shape}},
        enlarge y limits=0.05,
        enlarge x limits=0,
        width=\linewidth,
        height=0.4\linewidth,
    ]
        \addplot+ [thick, mark=none] table [x index=0, y index=1] {data/Q_norm_error_unique.dat};
        \addplot+ [thick, mark=none] table [x index=0, y index=1] {data/R_norm_error_unique.dat};
        \legend{$\left\Vert\hat{Q}(t) - Q\right\Vert$, $\left\Vert\hat{R}(t) - R\right\Vert$}
    \end{axis}
\end{tikzpicture} 
    \caption{A plot of the induced 2-norm of the error between the estimated $\hat{Q}$ (red) and $\hat{R}$ (blue) matrices and the expert's $Q$ and $R$ matrices as a function of time for the example that admits a unique solution.}
    \label{fig:Toy_QR_Diff_uni}
\end{figure}
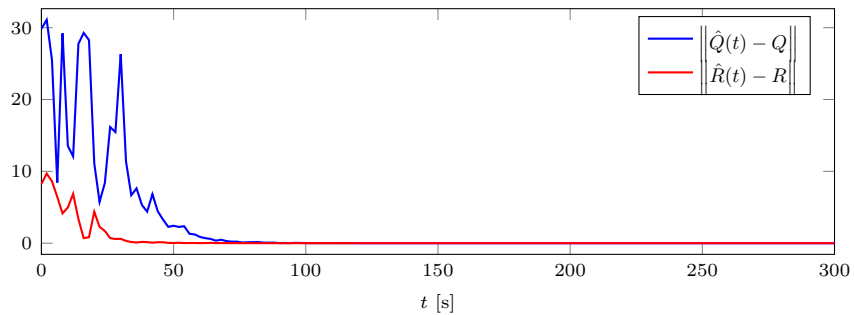

If the system matrix for the system in Section \ref{Section: Non-Unique} is changed to
\begin{equation*}
    A = \begin{bmatrix}
            1 & 0.4 & 1.6\\
            3.7 & 1.6 & -3.1\\
            -3.2 & 0.4 & 4.6
        \end{bmatrix},
\end{equation*}
then the state space model no longer admits a product structure and the corresponding IRL problem admits a unique solution. Fig. \ref{fig:Toy_QR_Diff_uni} indicates that when the IRL problem has a unique solution, the HSO developed in this paper recovers the true cost functional. As such, the HSO developed here is a proper extension of the HSO in \cite{SCC.Self.Coleman.ea2021a}.

\subsection{Kalman gain and the effects of measurement noise}\label{Section: Non-Unique with Noise}
This simulation provides insight into the noise robustness of the RHSO and its Kalman filter implementation (RHSO-KF). This investigation is purely heuristic in nature as the analysis does not consider noise. In RHSO-KF, the matrices $K_3$ and $K_4\hat{\Sigma}^{\top}$ are replaced with two Kalman gains, one for estimation of $\hat{x}$ and another for estimation of $\hat{W}$, respectively. Zero-mean Gaussian noise is added to $y$ with three different noise variances, $R_1=\diag([0.01^2,0.01^2, 0.01^2])$, $R_2=\diag([0.1^2,0.1^2, 0.1^2])$, and $R_3=\diag([0.5^2,0.5^2, 0.5^2])$. Zero mean Gaussian noise is also added to the estimates $\hat{W}$, with covariance $50 I$. The process noise covariance matrix for the state and the parameters is set to $0.001 I$. Fifty Monte-Carlo simulations are conducted for each noise level. The same model and simulation setup as Section \ref{Section: Non-Unique} is used with the exception that the magnitude of the excitation signal is increased to $1$ for $R_3$.

To facilitate the comparison between the RHSO and the RHSO-KF, the estimated cost functionals are optimized by solving the corresponding linear-quadratic regulator problem to generate estimates of the optimal learner trajectories. The root-mean-square (RMS) value of the pointwise norm of the error between the learner's trajectories and the expert's trajectories is computed for each noise standard deviation (SD) in each trial. Fig. \ref{fig:TT_all} demonstrates the noise rejection advantage of the RHSO-KF using box plots that show the distribution of the RMS errors over the 50 trials the three noise levels for the the RHSO and the RHSO-KF.

Fig. \ref{fig:Kp_all} shows box plots that compare the steady-state RMS values (computed over the last 30 seconds of the simulation) of the pointwise induced 2-norm of the feedback gain estimation error obtained by the RHSO and the RHSO-KF. The results suggest that a Kalman gain can be used to reduce the error in the presence of noise.
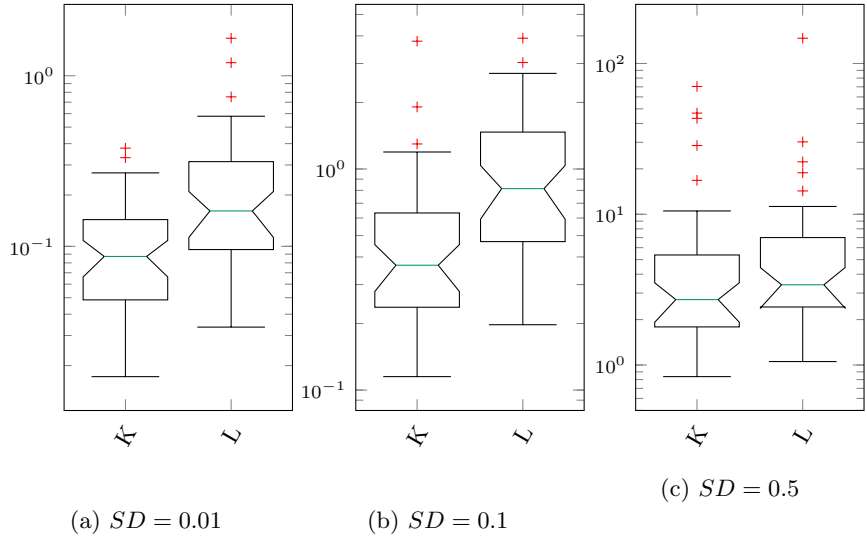
\begin{figure}
    \centering
    \begin{subfigure}[t]{.32\columnwidth}
        \centering
        \pgfplotstableread[col sep=comma,header=false]{data/Noise_TT_0.01.dat}\Table
        \pgfplotsset{
    boxplot/lower notch/.initial=\pgfutil@empty,
    boxplot/upper notch/.initial=\pgfutil@empty,
    boxplot/notch width/.initial=0.5,
    boxplot/draw/box/.code={%
        \draw[/pgfplots/boxplot/every box/.try]
        (boxplot box cs:\pgfplotsboxplotvalue{lower quartile},0)
        -- (boxplot box cs:\pgfplotsboxplotvalue{lower notch},0)
        -- (boxplot box cs:\pgfplotsboxplotvalue{median},0.5-\pgfplotsboxplotvalue{notch width}/2)
        -- (boxplot box cs:\pgfplotsboxplotvalue{upper notch},0)
        -- (boxplot box cs:\pgfplotsboxplotvalue{upper quartile},0)
        -- (boxplot box cs:\pgfplotsboxplotvalue{upper quartile},1)
        -- (boxplot box cs:\pgfplotsboxplotvalue{upper notch},1)
        -- (boxplot box cs:\pgfplotsboxplotvalue{median},0.5+\pgfplotsboxplotvalue{notch width}/2)
        -- (boxplot box cs:\pgfplotsboxplotvalue{lower notch},1)
        -- (boxplot box cs:\pgfplotsboxplotvalue{lower quartile},1)
        -- cycle;
    },%
    boxplot/draw/median/.code={%
        \draw[/pgfplots/boxplot/every median/.try, color=ForestGreen]
        (boxplot box cs:\pgfplotsboxplotvalue{median},0.5-\pgfplotsboxplotvalue{notch width}/2)
        -- (boxplot box cs:\pgfplotsboxplotvalue{median},0.5+\pgfplotsboxplotvalue{notch width}/2)
        ;
    },%
    boxplot prepared from table/.code={
        \def\tikz@plot@handler{\pgfplotsplothandlerboxplotprepared}%
        \pgfplotsset{
            /pgfplots/boxplot prepared from table/.cd,
            #1,
        }
    },
    /pgfplots/boxplot prepared from table/.cd,
    table/.code={\pgfplotstablecopy{#1}\to\boxplot@datatable},
    row/.initial=0,
    make style readable from table/.style={
        #1/.code={
            \pgfplotstablegetelem{\pgfkeysvalueof{/pgfplots/boxplot prepared from table/row}}{##1}\of\boxplot@datatable
            \pgfplotsset{boxplot/#1/.expand once={\pgfplotsretval}}
        }
    },
    make style readable from table=lower whisker,
    make style readable from table=upper whisker,
    make style readable from table=lower quartile,
    make style readable from table=upper quartile,
    make style readable from table=median,
    make style readable from table=lower notch,
    make style readable from table=upper notch
}
\begin{tikzpicture}
    \pgfplotstablegetrowsof{\Table}
    \pgfmathsetmacro\lastRowIndx{\pgfplotsretval-1}
    \pgfplotstablegetcolsof{\Table}
    \pgfmathsetmacro\lastcolIndx{\pgfplotsretval-1}
    \edef\lastCol{\pgfplotsretval}
    \ifnum \lastCol > 8
        \def\outlierList{8}
        \pgfplotsinvokeforeach{9,...,\lastcolIndx}{\edef\outlierList{\outlierList,#1}}
        \pgfplotstabletranspose[columns/.expanded=\outlierList]{\Outliers}{\Table}
    \fi
    \begin{axis}[
        width = 1.19\columnwidth,
        height = 1.8\columnwidth,
        label style={font=\scriptsize},
        tick label style={font=\scriptsize},
        boxplot/draw direction=y,
        ymode = log,
        xtick = {0,...,\lastRowIndx},
        xticklabel style = {align=center, font=\small, rotate=60},
        xticklabels from table={\Table}{[index] 0}
        ]
        \pgfplotsinvokeforeach {0,...,\lastRowIndx} {
            \ifnum \lastCol > 8
                \addplot[
                    mark=+,
                    mark options={red},
                    boxplot prepared from table={
                        table=\Table,
                        row=#1,
                        lower whisker=1,
                        upper whisker=5,
                        lower quartile=2,
                        upper quartile=4,
                        lower notch=6,
                        upper notch=7,
                        median=3,
                    },
                    boxplot prepared={draw position=#1}
                ] table [y=#1] {\Outliers};
            \else
                \addplot[
                    mark=+,
                    mark options={red},
                    boxplot prepared from table={
                        table=\Table,
                        row=#1,
                        lower whisker=1,
                        upper whisker=5,
                        lower quartile=2,
                        upper quartile=4,
                        lower notch=6,
                        upper notch=7,
                        median=3,
                    },
                    boxplot prepared={draw position=#1}
                ] coordinates {};
            \fi
        }
    \end{axis}
\end{tikzpicture}
        \caption{$SD=0.01$}
        \label{fig:sub4}
    \end{subfigure}%
    \begin{subfigure}[t]{.32\columnwidth}
        \centering
        \pgfplotstableread[col sep=comma,header=false]{data/Noise_TT_0.1.dat}\Table
        \pgfplotsset{
    boxplot/lower notch/.initial=\pgfutil@empty,
    boxplot/upper notch/.initial=\pgfutil@empty,
    boxplot/notch width/.initial=0.5,
    boxplot/draw/box/.code={%
        \draw[/pgfplots/boxplot/every box/.try]
        (boxplot box cs:\pgfplotsboxplotvalue{lower quartile},0)
        -- (boxplot box cs:\pgfplotsboxplotvalue{lower notch},0)
        -- (boxplot box cs:\pgfplotsboxplotvalue{median},0.5-\pgfplotsboxplotvalue{notch width}/2)
        -- (boxplot box cs:\pgfplotsboxplotvalue{upper notch},0)
        -- (boxplot box cs:\pgfplotsboxplotvalue{upper quartile},0)
        -- (boxplot box cs:\pgfplotsboxplotvalue{upper quartile},1)
        -- (boxplot box cs:\pgfplotsboxplotvalue{upper notch},1)
        -- (boxplot box cs:\pgfplotsboxplotvalue{median},0.5+\pgfplotsboxplotvalue{notch width}/2)
        -- (boxplot box cs:\pgfplotsboxplotvalue{lower notch},1)
        -- (boxplot box cs:\pgfplotsboxplotvalue{lower quartile},1)
        -- cycle;
    },%
    boxplot/draw/median/.code={%
        \draw[/pgfplots/boxplot/every median/.try, color=ForestGreen]
        (boxplot box cs:\pgfplotsboxplotvalue{median},0.5-\pgfplotsboxplotvalue{notch width}/2)
        -- (boxplot box cs:\pgfplotsboxplotvalue{median},0.5+\pgfplotsboxplotvalue{notch width}/2)
        ;
    },%
    boxplot prepared from table/.code={
        \def\tikz@plot@handler{\pgfplotsplothandlerboxplotprepared}%
        \pgfplotsset{
            /pgfplots/boxplot prepared from table/.cd,
            #1,
        }
    },
    /pgfplots/boxplot prepared from table/.cd,
    table/.code={\pgfplotstablecopy{#1}\to\boxplot@datatable},
    row/.initial=0,
    make style readable from table/.style={
        #1/.code={
            \pgfplotstablegetelem{\pgfkeysvalueof{/pgfplots/boxplot prepared from table/row}}{##1}\of\boxplot@datatable
            \pgfplotsset{boxplot/#1/.expand once={\pgfplotsretval}}
        }
    },
    make style readable from table=lower whisker,
    make style readable from table=upper whisker,
    make style readable from table=lower quartile,
    make style readable from table=upper quartile,
    make style readable from table=median,
    make style readable from table=lower notch,
    make style readable from table=upper notch
}
\begin{tikzpicture}
    \pgfplotstablegetrowsof{\Table}
    \pgfmathsetmacro\lastRowIndx{\pgfplotsretval-1}
    \pgfplotstablegetcolsof{\Table}
    \pgfmathsetmacro\lastcolIndx{\pgfplotsretval-1}
    \edef\lastCol{\pgfplotsretval}
    \ifnum \lastCol > 8
        \def\outlierList{8}
        \pgfplotsinvokeforeach{9,...,\lastcolIndx}{\edef\outlierList{\outlierList,#1}}
        \pgfplotstabletranspose[columns/.expanded=\outlierList]{\Outliers}{\Table}
    \fi
    \begin{axis}[
        width = 1.19\columnwidth,
        height = 1.8\columnwidth,
        label style={font=\scriptsize},
        tick label style={font=\scriptsize},
        boxplot/draw direction=y,
        ymode = log,
        xtick = {0,...,\lastRowIndx},
        xticklabel style = {align=center, font=\small, rotate=60},
        xticklabels from table={\Table}{[index] 0}
        ]
        \pgfplotsinvokeforeach {0,...,\lastRowIndx} {
            \ifnum \lastCol > 8
                \addplot[
                    mark=+,
                    mark options={red},
                    boxplot prepared from table={
                        table=\Table,
                        row=#1,
                        lower whisker=1,
                        upper whisker=5,
                        lower quartile=2,
                        upper quartile=4,
                        lower notch=6,
                        upper notch=7,
                        median=3,
                    },
                    boxplot prepared={draw position=#1}
                ] table [y=#1] {\Outliers};
            \else
                \addplot[
                    mark=+,
                    mark options={red},
                    boxplot prepared from table={
                        table=\Table,
                        row=#1,
                        lower whisker=1,
                        upper whisker=5,
                        lower quartile=2,
                        upper quartile=4,
                        lower notch=6,
                        upper notch=7,
                        median=3,
                    },
                    boxplot prepared={draw position=#1}
                ] coordinates {};
            \fi
        }
    \end{axis}
\end{tikzpicture}
        \caption{$SD=0.1$}
        \label{fig:sub5}
    \end{subfigure}%
    \begin{subfigure}[t]{.32\columnwidth}
        \centering
        \pgfplotstableread[col sep=comma,header=false]{data/Noise_TT_0.5.dat}\Table
        \pgfplotsset{
    boxplot/lower notch/.initial=\pgfutil@empty,
    boxplot/upper notch/.initial=\pgfutil@empty,
    boxplot/notch width/.initial=0.5,
    boxplot/draw/box/.code={%
        \draw[/pgfplots/boxplot/every box/.try]
        (boxplot box cs:\pgfplotsboxplotvalue{lower quartile},0)
        -- (boxplot box cs:\pgfplotsboxplotvalue{lower notch},0)
        -- (boxplot box cs:\pgfplotsboxplotvalue{median},0.5-\pgfplotsboxplotvalue{notch width}/2)
        -- (boxplot box cs:\pgfplotsboxplotvalue{upper notch},0)
        -- (boxplot box cs:\pgfplotsboxplotvalue{upper quartile},0)
        -- (boxplot box cs:\pgfplotsboxplotvalue{upper quartile},1)
        -- (boxplot box cs:\pgfplotsboxplotvalue{upper notch},1)
        -- (boxplot box cs:\pgfplotsboxplotvalue{median},0.5+\pgfplotsboxplotvalue{notch width}/2)
        -- (boxplot box cs:\pgfplotsboxplotvalue{lower notch},1)
        -- (boxplot box cs:\pgfplotsboxplotvalue{lower quartile},1)
        -- cycle;
    },%
    boxplot/draw/median/.code={%
        \draw[/pgfplots/boxplot/every median/.try, color=ForestGreen]
        (boxplot box cs:\pgfplotsboxplotvalue{median},0.5-\pgfplotsboxplotvalue{notch width}/2)
        -- (boxplot box cs:\pgfplotsboxplotvalue{median},0.5+\pgfplotsboxplotvalue{notch width}/2)
        ;
    },%
    boxplot prepared from table/.code={
        \def\tikz@plot@handler{\pgfplotsplothandlerboxplotprepared}%
        \pgfplotsset{
            /pgfplots/boxplot prepared from table/.cd,
            #1,
        }
    },
    /pgfplots/boxplot prepared from table/.cd,
    table/.code={\pgfplotstablecopy{#1}\to\boxplot@datatable},
    row/.initial=0,
    make style readable from table/.style={
        #1/.code={
            \pgfplotstablegetelem{\pgfkeysvalueof{/pgfplots/boxplot prepared from table/row}}{##1}\of\boxplot@datatable
            \pgfplotsset{boxplot/#1/.expand once={\pgfplotsretval}}
        }
    },
    make style readable from table=lower whisker,
    make style readable from table=upper whisker,
    make style readable from table=lower quartile,
    make style readable from table=upper quartile,
    make style readable from table=median,
    make style readable from table=lower notch,
    make style readable from table=upper notch
}
\begin{tikzpicture}
    \pgfplotstablegetrowsof{\Table}
    \pgfmathsetmacro\lastRowIndx{\pgfplotsretval-1}
    \pgfplotstablegetcolsof{\Table}
    \pgfmathsetmacro\lastcolIndx{\pgfplotsretval-1}
    \edef\lastCol{\pgfplotsretval}
    \ifnum \lastCol > 8
        \def\outlierList{8}
        \pgfplotsinvokeforeach{9,...,\lastcolIndx}{\edef\outlierList{\outlierList,#1}}
        \pgfplotstabletranspose[columns/.expanded=\outlierList]{\Outliers}{\Table}
    \fi
    \begin{axis}[
        width = 1.19\columnwidth,
        height = 1.8\columnwidth,
        label style={font=\scriptsize},
        tick label style={font=\scriptsize},
        boxplot/draw direction=y,
        ymode = log,
        xtick = {0,...,\lastRowIndx},
        xticklabel style = {align=center, font=\small, rotate=60},
        xticklabels from table={\Table}{[index] 0}
        ]
        \pgfplotsinvokeforeach {0,...,\lastRowIndx} {
            \ifnum \lastCol > 8
                \addplot[
                    mark=+,
                    mark options={red},
                    boxplot prepared from table={
                        table=\Table,
                        row=#1,
                        lower whisker=1,
                        upper whisker=5,
                        lower quartile=2,
                        upper quartile=4,
                        lower notch=6,
                        upper notch=7,
                        median=3,
                    },
                    boxplot prepared={draw position=#1}
                ] table [y=#1] {\Outliers};
            \else
                \addplot[
                    mark=+,
                    mark options={red},
                    boxplot prepared from table={
                        table=\Table,
                        row=#1,
                        lower whisker=1,
                        upper whisker=5,
                        lower quartile=2,
                        upper quartile=4,
                        lower notch=6,
                        upper notch=7,
                        median=3,
                    },
                    boxplot prepared={draw position=#1}
                ] coordinates {};
            \fi
        }
    \end{axis}
\end{tikzpicture}
        \caption{$SD=0.5$}
        \label{fig:sub6}
    \end{subfigure}
    \caption{Box plot of the RMS error between the expert's trajectory and the learner's trajectory, generated by optimizing the learner's estimated cost functional. The three subplots correspond to the three noise levels and the labels K and L correspond to the RHSO-KF and the RHSO, respectively.}
    \label{fig:TT_all}
\end{figure}
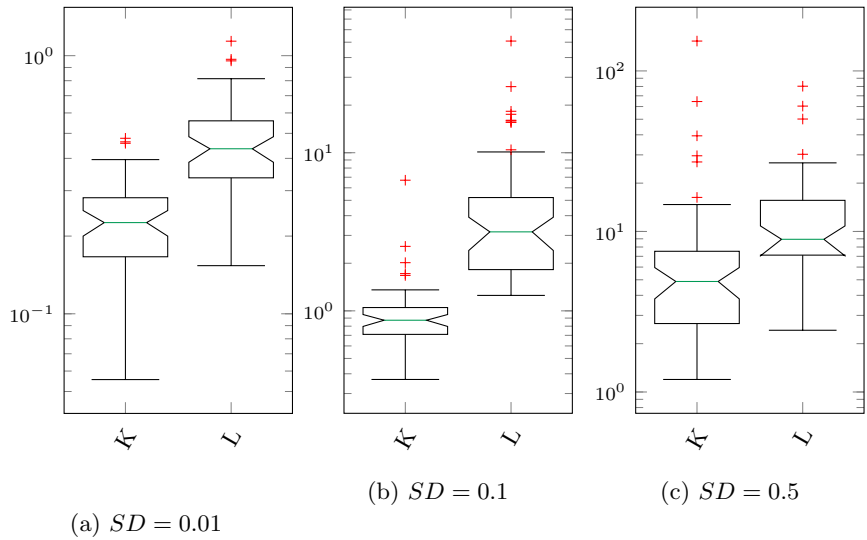
\begin{figure}
   \centering
   \begin{subfigure}[t]{.32\columnwidth}
       \centering
       \pgfplotstableread[col sep=comma,header=false]{data/Noise_Kp_0.01.dat}\Table
       \pgfplotsset{
    boxplot/lower notch/.initial=\pgfutil@empty,
    boxplot/upper notch/.initial=\pgfutil@empty,
    boxplot/notch width/.initial=0.5,
    boxplot/draw/box/.code={%
        \draw[/pgfplots/boxplot/every box/.try]
        (boxplot box cs:\pgfplotsboxplotvalue{lower quartile},0)
        -- (boxplot box cs:\pgfplotsboxplotvalue{lower notch},0)
        -- (boxplot box cs:\pgfplotsboxplotvalue{median},0.5-\pgfplotsboxplotvalue{notch width}/2)
        -- (boxplot box cs:\pgfplotsboxplotvalue{upper notch},0)
        -- (boxplot box cs:\pgfplotsboxplotvalue{upper quartile},0)
        -- (boxplot box cs:\pgfplotsboxplotvalue{upper quartile},1)
        -- (boxplot box cs:\pgfplotsboxplotvalue{upper notch},1)
        -- (boxplot box cs:\pgfplotsboxplotvalue{median},0.5+\pgfplotsboxplotvalue{notch width}/2)
        -- (boxplot box cs:\pgfplotsboxplotvalue{lower notch},1)
        -- (boxplot box cs:\pgfplotsboxplotvalue{lower quartile},1)
        -- cycle;
    },%
    boxplot/draw/median/.code={%
        \draw[/pgfplots/boxplot/every median/.try, color=ForestGreen]
        (boxplot box cs:\pgfplotsboxplotvalue{median},0.5-\pgfplotsboxplotvalue{notch width}/2)
        -- (boxplot box cs:\pgfplotsboxplotvalue{median},0.5+\pgfplotsboxplotvalue{notch width}/2)
        ;
    },%
    boxplot prepared from table/.code={
        \def\tikz@plot@handler{\pgfplotsplothandlerboxplotprepared}%
        \pgfplotsset{
            /pgfplots/boxplot prepared from table/.cd,
            #1,
        }
    },
    /pgfplots/boxplot prepared from table/.cd,
    table/.code={\pgfplotstablecopy{#1}\to\boxplot@datatable},
    row/.initial=0,
    make style readable from table/.style={
        #1/.code={
            \pgfplotstablegetelem{\pgfkeysvalueof{/pgfplots/boxplot prepared from table/row}}{##1}\of\boxplot@datatable
            \pgfplotsset{boxplot/#1/.expand once={\pgfplotsretval}}
        }
    },
    make style readable from table=lower whisker,
    make style readable from table=upper whisker,
    make style readable from table=lower quartile,
    make style readable from table=upper quartile,
    make style readable from table=median,
    make style readable from table=lower notch,
    make style readable from table=upper notch
}
\begin{tikzpicture}
    \pgfplotstablegetrowsof{\Table}
    \pgfmathsetmacro\lastRowIndx{\pgfplotsretval-1}
    \pgfplotstablegetcolsof{\Table}
    \pgfmathsetmacro\lastcolIndx{\pgfplotsretval-1}
    \edef\lastCol{\pgfplotsretval}
    \ifnum \lastCol > 8
        \def\outlierList{8}
        \pgfplotsinvokeforeach{9,...,\lastcolIndx}{\edef\outlierList{\outlierList,#1}}
        \pgfplotstabletranspose[columns/.expanded=\outlierList]{\Outliers}{\Table}
    \fi
    \begin{axis}[
        width = 1.19\columnwidth,
        height = 1.8\columnwidth,
        label style={font=\scriptsize},
        tick label style={font=\scriptsize},
        boxplot/draw direction=y,
        ymode = log,
        xtick = {0,...,\lastRowIndx},
        xticklabel style = {align=center, font=\small, rotate=60},
        xticklabels from table={\Table}{[index] 0}
        ]
        \pgfplotsinvokeforeach {0,...,\lastRowIndx} {
            \ifnum \lastCol > 8
                \addplot[
                    mark=+,
                    mark options={red},
                    boxplot prepared from table={
                        table=\Table,
                        row=#1,
                        lower whisker=1,
                        upper whisker=5,
                        lower quartile=2,
                        upper quartile=4,
                        lower notch=6,
                        upper notch=7,
                        median=3,
                    },
                    boxplot prepared={draw position=#1}
                ] table [y=#1] {\Outliers};
            \else
                \addplot[
                    mark=+,
                    mark options={red},
                    boxplot prepared from table={
                        table=\Table,
                        row=#1,
                        lower whisker=1,
                        upper whisker=5,
                        lower quartile=2,
                        upper quartile=4,
                        lower notch=6,
                        upper notch=7,
                        median=3,
                    },
                    boxplot prepared={draw position=#1}
                ] coordinates {};
            \fi
        }
    \end{axis}
\end{tikzpicture}
       \caption{$SD=0.01$}
       \label{fig:sub1}
   \end{subfigure}%
   \begin{subfigure}[t]{.32\columnwidth}
       \centering
       \pgfplotstableread[col sep=comma,header=false]{data/Noise_Kp_0.1.dat}\Table
       \pgfplotsset{
    boxplot/lower notch/.initial=\pgfutil@empty,
    boxplot/upper notch/.initial=\pgfutil@empty,
    boxplot/notch width/.initial=0.5,
    boxplot/draw/box/.code={%
        \draw[/pgfplots/boxplot/every box/.try]
        (boxplot box cs:\pgfplotsboxplotvalue{lower quartile},0)
        -- (boxplot box cs:\pgfplotsboxplotvalue{lower notch},0)
        -- (boxplot box cs:\pgfplotsboxplotvalue{median},0.5-\pgfplotsboxplotvalue{notch width}/2)
        -- (boxplot box cs:\pgfplotsboxplotvalue{upper notch},0)
        -- (boxplot box cs:\pgfplotsboxplotvalue{upper quartile},0)
        -- (boxplot box cs:\pgfplotsboxplotvalue{upper quartile},1)
        -- (boxplot box cs:\pgfplotsboxplotvalue{upper notch},1)
        -- (boxplot box cs:\pgfplotsboxplotvalue{median},0.5+\pgfplotsboxplotvalue{notch width}/2)
        -- (boxplot box cs:\pgfplotsboxplotvalue{lower notch},1)
        -- (boxplot box cs:\pgfplotsboxplotvalue{lower quartile},1)
        -- cycle;
    },%
    boxplot/draw/median/.code={%
        \draw[/pgfplots/boxplot/every median/.try, color=ForestGreen]
        (boxplot box cs:\pgfplotsboxplotvalue{median},0.5-\pgfplotsboxplotvalue{notch width}/2)
        -- (boxplot box cs:\pgfplotsboxplotvalue{median},0.5+\pgfplotsboxplotvalue{notch width}/2)
        ;
    },%
    boxplot prepared from table/.code={
        \def\tikz@plot@handler{\pgfplotsplothandlerboxplotprepared}%
        \pgfplotsset{
            /pgfplots/boxplot prepared from table/.cd,
            #1,
        }
    },
    /pgfplots/boxplot prepared from table/.cd,
    table/.code={\pgfplotstablecopy{#1}\to\boxplot@datatable},
    row/.initial=0,
    make style readable from table/.style={
        #1/.code={
            \pgfplotstablegetelem{\pgfkeysvalueof{/pgfplots/boxplot prepared from table/row}}{##1}\of\boxplot@datatable
            \pgfplotsset{boxplot/#1/.expand once={\pgfplotsretval}}
        }
    },
    make style readable from table=lower whisker,
    make style readable from table=upper whisker,
    make style readable from table=lower quartile,
    make style readable from table=upper quartile,
    make style readable from table=median,
    make style readable from table=lower notch,
    make style readable from table=upper notch
}
\begin{tikzpicture}
    \pgfplotstablegetrowsof{\Table}
    \pgfmathsetmacro\lastRowIndx{\pgfplotsretval-1}
    \pgfplotstablegetcolsof{\Table}
    \pgfmathsetmacro\lastcolIndx{\pgfplotsretval-1}
    \edef\lastCol{\pgfplotsretval}
    \ifnum \lastCol > 8
        \def\outlierList{8}
        \pgfplotsinvokeforeach{9,...,\lastcolIndx}{\edef\outlierList{\outlierList,#1}}
        \pgfplotstabletranspose[columns/.expanded=\outlierList]{\Outliers}{\Table}
    \fi
    \begin{axis}[
        width = 1.19\columnwidth,
        height = 1.8\columnwidth,
        label style={font=\scriptsize},
        tick label style={font=\scriptsize},
        boxplot/draw direction=y,
        ymode = log,
        xtick = {0,...,\lastRowIndx},
        xticklabel style = {align=center, font=\small, rotate=60},
        xticklabels from table={\Table}{[index] 0}
        ]
        \pgfplotsinvokeforeach {0,...,\lastRowIndx} {
            \ifnum \lastCol > 8
                \addplot[
                    mark=+,
                    mark options={red},
                    boxplot prepared from table={
                        table=\Table,
                        row=#1,
                        lower whisker=1,
                        upper whisker=5,
                        lower quartile=2,
                        upper quartile=4,
                        lower notch=6,
                        upper notch=7,
                        median=3,
                    },
                    boxplot prepared={draw position=#1}
                ] table [y=#1] {\Outliers};
            \else
                \addplot[
                    mark=+,
                    mark options={red},
                    boxplot prepared from table={
                        table=\Table,
                        row=#1,
                        lower whisker=1,
                        upper whisker=5,
                        lower quartile=2,
                        upper quartile=4,
                        lower notch=6,
                        upper notch=7,
                        median=3,
                    },
                    boxplot prepared={draw position=#1}
                ] coordinates {};
            \fi
        }
    \end{axis}
\end{tikzpicture}
       \caption{$SD=0.1$}
       \label{fig:sub2}
   \end{subfigure}%
   \begin{subfigure}[t]{.32\columnwidth}
       \centering
       \pgfplotstableread[col sep=comma,header=false]{data/Noise_Kp_0.5.dat}\Table
       \pgfplotsset{
    boxplot/lower notch/.initial=\pgfutil@empty,
    boxplot/upper notch/.initial=\pgfutil@empty,
    boxplot/notch width/.initial=0.5,
    boxplot/draw/box/.code={%
        \draw[/pgfplots/boxplot/every box/.try]
        (boxplot box cs:\pgfplotsboxplotvalue{lower quartile},0)
        -- (boxplot box cs:\pgfplotsboxplotvalue{lower notch},0)
        -- (boxplot box cs:\pgfplotsboxplotvalue{median},0.5-\pgfplotsboxplotvalue{notch width}/2)
        -- (boxplot box cs:\pgfplotsboxplotvalue{upper notch},0)
        -- (boxplot box cs:\pgfplotsboxplotvalue{upper quartile},0)
        -- (boxplot box cs:\pgfplotsboxplotvalue{upper quartile},1)
        -- (boxplot box cs:\pgfplotsboxplotvalue{upper notch},1)
        -- (boxplot box cs:\pgfplotsboxplotvalue{median},0.5+\pgfplotsboxplotvalue{notch width}/2)
        -- (boxplot box cs:\pgfplotsboxplotvalue{lower notch},1)
        -- (boxplot box cs:\pgfplotsboxplotvalue{lower quartile},1)
        -- cycle;
    },%
    boxplot/draw/median/.code={%
        \draw[/pgfplots/boxplot/every median/.try, color=ForestGreen]
        (boxplot box cs:\pgfplotsboxplotvalue{median},0.5-\pgfplotsboxplotvalue{notch width}/2)
        -- (boxplot box cs:\pgfplotsboxplotvalue{median},0.5+\pgfplotsboxplotvalue{notch width}/2)
        ;
    },%
    boxplot prepared from table/.code={
        \def\tikz@plot@handler{\pgfplotsplothandlerboxplotprepared}%
        \pgfplotsset{
            /pgfplots/boxplot prepared from table/.cd,
            #1,
        }
    },
    /pgfplots/boxplot prepared from table/.cd,
    table/.code={\pgfplotstablecopy{#1}\to\boxplot@datatable},
    row/.initial=0,
    make style readable from table/.style={
        #1/.code={
            \pgfplotstablegetelem{\pgfkeysvalueof{/pgfplots/boxplot prepared from table/row}}{##1}\of\boxplot@datatable
            \pgfplotsset{boxplot/#1/.expand once={\pgfplotsretval}}
        }
    },
    make style readable from table=lower whisker,
    make style readable from table=upper whisker,
    make style readable from table=lower quartile,
    make style readable from table=upper quartile,
    make style readable from table=median,
    make style readable from table=lower notch,
    make style readable from table=upper notch
}
\begin{tikzpicture}
    \pgfplotstablegetrowsof{\Table}
    \pgfmathsetmacro\lastRowIndx{\pgfplotsretval-1}
    \pgfplotstablegetcolsof{\Table}
    \pgfmathsetmacro\lastcolIndx{\pgfplotsretval-1}
    \edef\lastCol{\pgfplotsretval}
    \ifnum \lastCol > 8
        \def\outlierList{8}
        \pgfplotsinvokeforeach{9,...,\lastcolIndx}{\edef\outlierList{\outlierList,#1}}
        \pgfplotstabletranspose[columns/.expanded=\outlierList]{\Outliers}{\Table}
    \fi
    \begin{axis}[
        width = 1.19\columnwidth,
        height = 1.8\columnwidth,
        label style={font=\scriptsize},
        tick label style={font=\scriptsize},
        boxplot/draw direction=y,
        ymode = log,
        xtick = {0,...,\lastRowIndx},
        xticklabel style = {align=center, font=\small, rotate=60},
        xticklabels from table={\Table}{[index] 0}
        ]
        \pgfplotsinvokeforeach {0,...,\lastRowIndx} {
            \ifnum \lastCol > 8
                \addplot[
                    mark=+,
                    mark options={red},
                    boxplot prepared from table={
                        table=\Table,
                        row=#1,
                        lower whisker=1,
                        upper whisker=5,
                        lower quartile=2,
                        upper quartile=4,
                        lower notch=6,
                        upper notch=7,
                        median=3,
                    },
                    boxplot prepared={draw position=#1}
                ] table [y=#1] {\Outliers};
            \else
                \addplot[
                    mark=+,
                    mark options={red},
                    boxplot prepared from table={
                        table=\Table,
                        row=#1,
                        lower whisker=1,
                        upper whisker=5,
                        lower quartile=2,
                        upper quartile=4,
                        lower notch=6,
                        upper notch=7,
                        median=3,
                    },
                    boxplot prepared={draw position=#1}
                ] coordinates {};
            \fi
        }
    \end{axis}
\end{tikzpicture}
       \caption{$SD=0.5$}
       \label{fig:sub3}
   \end{subfigure}
   \caption{Box plot of the feedback gain estimation error. The three subplots correspond to three noise levels and the labels K and L correspond to the RHSO-KF and the RHSO, respectively.}
   \label{fig:Kp_all}
\end{figure}

\subsection{Discussion}
Each simulation shows the convergence of $\Delta$ to zero and the convergence of the estimated feedback matrix, $\hat{K}_P$, to the feedback matrix $K_{Ep}$ of the expert. In all simulations, the RHSO converges to either an equivalent solution or the true cost functional of the expert. Therefore, the RHSO is a complete extension to the HSO \cite{SCC.Self.Coleman.ea2021a} as it solves IRL problems with unique and non-unique solutions. The particular equivalent solution that the RHSO converges to depends on the initial estimates of the unknown weights $\hat{W}$.

As demonstrated by Fig. \ref{fig:Toy_QR_Diff_non}, convergence to an approximate equivalent solution is achieved in spite of failure to meet the FI condition throughout the simulation. The condition is met, however, at the end of the simulation. Fig. \ref{fig:Toy_QR_Diff_non} thus indicates that the FI condition is sufficient but not necessary for the RHSO to converge to approximate equivalent solutions. When $K_4$ is selected as $(\hat{\Sigma}^{\top}\hat{\Sigma}+\epsilon I)^{-1}$, $\Delta$ converges to zero and either a unique or an equivalent solution is obtained, regardless of the magnitude of $\epsilon$. This result is at odds with regularization used in ridge regression, where convergence with an $\epsilon-$dependent bound is obtained. Especially interesting is the fact that offline ridge regression \cite{SCC.Tibshirani1996} using matrices $\Sigma_u$ and $\hat{\Sigma}$ that contain all of the available data fail at finding a $\hat{W}$ that constitutes an equivalent solution to the IRL problem.

\section{Conclusion}\label{Section: Conclusion}
In this paper, a novel framework for the estimation of a cost functional is developed for IRL problems with multiple solutions. The developed technique is a modification of the HSO in \cite{SCC.Self.Coleman.ea2021a}. This modification, while simple, requires a novel analysis approach. The analysis reveals new data-informativity conditions required for convergence of the update laws to an equivalent solution when multiple solutions are present. It is further shown that the RHSO is a proper extension of the HSO, in the sense that it converges to the true cost functional of the expert when the IRL problem has a unique solution.

Simulations demonstrate that the developed adaptive update laws are able to converge to equivalent solutions in IRL problems where offline ridge-regression fails to generate useful solutions. 
While theoretical analysis of the case with sensor noise is a part of future research, the Monte-Carlo simulations demonstrate that in the presence of measurement noise, the RHSO can be implemented using a Kalman gain instead of a Luenberger-like gain for improved performance. Future research will include applications of the developed method to real-world problems such as learning the cost function of pilots flying unmanned air vehicles using input-output measurements.
\bibliographystyle{plain}
\bibliography{scc,sccmaster,scctemp}
\end{document}